\documentclass{article}
\usepackage{amssymb}
\usepackage{amsfonts}
\usepackage{graphicx}
\usepackage{amsmath}
\usepackage[margin=1in]{geometry}
\usepackage{scalefnt}
\usepackage{ulem}
\usepackage{hyperref}
\usepackage{indentfirst}

\numberwithin{equation}{section}
\input{tcilatex}

\begin{document}

\title{Fedosov Observables on Constant Curvature Manifolds and the
Klein-Gordon Equation}
\author{Philip C. Tillman$^{1}$, George A.J. Sparling$^{2}$ \\
%EndAName
$^{1}$Department of Physics and Astronomy, University of Pittsburgh,
Pittsburgh, PA, USA\\
$^{2}$Department of Mathematics, University of Pittsburgh, Pittsburgh, PA,
USA\\
email:$^{1}$phil.tillman@gmail.com $^{2}$sparling@twistor.org}
\date{\today }
\maketitle

\begin{abstract}
In this paper we construct the Fedosov star-algebra of observables on the
phase-space of a single particle in the case of all (finite-dimensional)
constant curvature manifolds imbeddable in a flat space with codimension 1.
This set of spaces includes the two-sphere and de Sitter (dS)/anti-de Sitter
(AdS) space-times. The algebra of observables was constructed by DQ
techniques using, in particular, the algorithm provided by Fedosov.

The purpose of this paper was three-fold. One was to verify that DQ gave the
same results as previous analyses of these spaces. Another was to verify
that the formal series used in the conventional treatment converged by
obtaining exact and nonperturbative results for these spaces. The last was
to further develop and understand the technology of the Fedosov algorithm.
\end{abstract}

\section{Introduction}

Deformation quantization (DQ) yields an equivalent mathematical formulation
of quantum mechanics on phase-space. The key difference between DQ and an
operator formulation is that in DQ observables from classical theories are
not mapped to operators--they simply stay the same. What does change or,
more accurately, is introduced is a funny thing called a star-product (see 
%TCIMACRO{%
%\TeXButton{Hirshfeld A. and Henselder P. 2002a}{\hyperlink{ref1}{Hirshfeld A. and Henselder P. 2002a}}}%
%BeginExpansion
\hyperlink{ref1}{Hirshfeld A. and Henselder P. 2002a}%
%EndExpansion
, and 
%TCIMACRO{%
%\TeXButton{Hancock J. et al 2004}{\hyperlink{ref1}{Hancock J. \textit{et al} 2004}}}%
%BeginExpansion
\hyperlink{ref1}{Hancock J. \textit{et al} 2004}%
%EndExpansion
).

The star product is simply a map $\ast $ that maps two functions on
phase-space to another in a way that can reproduce quantum mechanics. In
other words, the resulting star-algebra is isomorphic to the space of linear
operators on a Hilbert space which is the usual observable algebra one works
with in quantum mechanics. Key relations in flat space like:%
\begin{equation*}
\left[ \hat{x}^{\mu },\hat{x}^{\nu }\right] =0~~~,~~~\left[ \hat{x}^{\mu },%
\hat{p}_{\nu }\right] =i\hbar \delta _{\nu }^{\mu }~~,~~~\left[ \hat{p}_{\mu
},\hat{p}_{\nu }\right] =0
\end{equation*}%
are reproduced in the star-algebra as:%
\begin{equation*}
\left[ x^{\mu },x^{\nu }\right] _{\ast }=0~~~,~~~\left[ x^{\mu },p_{\nu }%
\right] _{\ast }=i\hbar \delta _{\nu }^{\mu }~~~,~~~\left[ p_{\mu },p_{\nu }%
\right] _{\ast }=0
\end{equation*}%
where the commutator $\left[ f,g\right] _{\ast }=f\ast g-g\ast f$ for all
phase-space functions $f$ and $g$. Also, the star-product is associative and
linear as is dictated by quantum mechanics and the presence of Hilbert space
representations.

%TCIMACRO{%
%\TeXButton{Fedosov B. (1996)}{\hyperlink{ref1}{Fedosov B. (1996)}} }%
%BeginExpansion
\hyperlink{ref1}{Fedosov B. (1996)}
%EndExpansion
has provided an algorithm to construct a star-product as a formal series in $%
\hbar $ on any finite-dimensional symplectic manifold. The algorithm's power
is that it is geometrical and does not rely on coordinate dependent things.
To understand how the Fedosov the basic idea of the algorithm we should
understand the Groenewold-Moyal star-product.\pagebreak

The birth of Moyal star-product (hence the birth of DQ) relys on the
quantization map given by the Weyl quantization map (usually written as an
integral transform) $\mathcal{W}$. The Weyl quantization map assigns to each
phase-space function a unique observable by symmetric ordering, for example:%
\begin{equation*}
\mathcal{W}\left( x^{2}p\right) =\frac{1}{3}\left( \hat{x}^{2}\hat{p}+\hat{x}%
\hat{p}\hat{x}+\hat{p}\hat{x}^{2}\right)
\end{equation*}%
in general we have $\mathcal{W}\left( ax+bp\right) ^{n}=\left( a\hat{x}+b%
\hat{p}\right) ^{n}$. Now, we can use Wigner's inverse map $\mathcal{W}^{-1}$
(the inverse of the integral transform) and we can find the Groenewold-Moyal
star-product defined as:%
\begin{equation*}
f\ast g:=\mathcal{W}^{-1}\left( \mathcal{W}\left( f\right) \mathcal{W}\left(
g\right) \right)
\end{equation*}%
%TCIMACRO{%
%\TeXButton{Groenewold H. (1946)}{\hyperlink{ref1}{Groenewold H. (1946)}} }%
%BeginExpansion
\hyperlink{ref1}{Groenewold H. (1946)}
%EndExpansion
(and later 
%TCIMACRO{\TeXButton{Moyal J. 1949}{\hyperlink{ref1}{Moyal J. 1949}}}%
%BeginExpansion
\hyperlink{ref1}{Moyal J. 1949}%
%EndExpansion
) investigated this formula and found a remarkable result:%
\begin{equation}
f\ast g=f\exp \left[ \frac{i\hbar }{2}\left( \frac{\overleftarrow{\partial }%
}{\partial x^{\mu }}\frac{\overrightarrow{\partial }}{\partial p_{\mu }}-%
\frac{\overleftarrow{\partial }}{\partial p_{\mu }}\frac{\overrightarrow{%
\partial }}{\partial x^{\mu }}\right) \right] g  \notag
\end{equation}%
In a coordinate independent formulation we have:%
\begin{eqnarray}
f\ast g &=&f\exp \left[ \left( i\hbar /2\right) \overleftrightarrow{P}\right]
g  \label{Moyal} \\
&=&\sum_{A,B,j}^{\infty }\left( i\hbar /2\right) ^{j}\omega
^{A_{1}B_{1}}\cdots \omega ^{A_{j}B_{j}}/j!(\partial _{A_{1}}\cdots \partial
_{A_{j}}f)(\partial _{B_{1}}\cdots \partial _{B_{j}}g)  \notag
\end{eqnarray}%
\begin{equation*}
\overleftrightarrow{P}:=\overleftarrow{\partial }_{A}\omega _{AB}%
\overrightarrow{\partial }_{B}
\end{equation*}%
where $\overleftrightarrow{P}$ is the Poisson bracket and $\partial _{A}$ is
a (flat) torsion-free phase-space connection ($\partial \otimes \omega =0$).
Also, $q^{A}=\left( x^{\mu },p_{\mu }\right) $ and $\partial
_{A}q^{B}=\delta _{A}^{B}$. The capital Latin indices $A$, $B$, etc. are
numerical phase-space indices and run from $1$ to $2n$ while the Greek
lowercase indices represent numerical space-time indices. We will sometimes
use abstract space-time indices represented by lowercase Latin letters.

The Fedosov algorithm does the same basic thing first find the quantization
map he calls $\sigma ^{-1}$ then use its inverse to define the Fedosov
star-product by:%
\begin{equation*}
f\ast g:=\sigma \left( \sigma ^{-1}\left( f\right) \sigma ^{-1}\left(
g\right) \right)
\end{equation*}

In this paper we construct the Fedosov star-product on an arbitrary
(finite-dimensional) constant curvature manifold of codimension one (the
precise definition of this manifold will be given later) by constructing the
map $\sigma ^{-1}$. This paper is a straightforward generalization of our
previous paper 
%TCIMACRO{%
%\TeXButton{Tillman P. and Sparling G. (2006)}{\hyperlink{ref1}{Tillman P. and Sparling G. (2006)}}}%
%BeginExpansion
\hyperlink{ref1}{Tillman P. and Sparling G. (2006)}%
%EndExpansion
, where in that paper we considered the two-sphere case of which the case
considered now subsumes. Along the way we will derive some formulas (and
some properties thereof) that are completely general for a
finite-dimensional phase-space of a configuration space which represents our
space-time. We feel that they may useful for future calculations of the
Fedosov star-product.

What is shown is that following an \textit{exact and nonperturbative}
calculation, the resulting star-algebra is the pseudo-orthogonal group $%
\mathbb{SO}\left( p+1,q+1\right) $ where the $p$ and the $q$ are fixed by
the embedding formula. This also addresses the question of convergence of
the map $\sigma ^{-1}$ which is a critical problem of the general Fedosov
star and DQ in general. Also, the Klein-Gordon equation is given by a
Casimir invariant of a subgroup, either $\mathbb{SO}\left( p,q+1\right) $ or 
$\mathbb{SO}\left( p+1,q\right) $ (the choice again depends on the embedding
formula). We note that the subgroup $\mathbb{SO}\left( p,q+1\right) $ or $%
\mathbb{SO}\left( p+1,q\right) $ is the symmetry group of this constant
curvature manifold. These results are completely expected and consistent
with the analysis of 
%TCIMACRO{%
%\TeXButton{Fr\o nsdal C. (1965, 1973, 1975a, 1975b)}{\hyperlink{ref1}{Fr\o nsdal C. (1965, 1973, 1975a, 1975b)}} }%
%BeginExpansion
\hyperlink{ref1}{Fr\o nsdal C. (1965, 1973, 1975a, 1975b)}
%EndExpansion
which is the standard theory of particles on de Sitter (dS) and anti-de
Sitter (AdS) space-times in 1+3 dimensions. The advantage of our result
using the Fedosov star-product is that it is algorithmic and whereas the
results achieved by Fr\o nsdal and others rely, crucially, on symmetries of
the particular case considered.\pagebreak 

\subsection{Outline}

Before the reader begins this paper, they should familiarize themselves with
the notations in 
%TCIMACRO{\TeXButton{appendix A}{\hyperlink{Appendix A}{appendix A}}}%
%BeginExpansion
\hyperlink{Appendix A}{appendix A}%
%EndExpansion
. In 
%TCIMACRO{%
%\TeXButton{The Fedosov Star-Product}{\hyperref[s:The Fedosov Star-Product]{section~\ref*{s:The Fedosov Star-Product}}} }%
%BeginExpansion
\hyperref[s:The Fedosov Star-Product]{section~\ref*{s:The Fedosov Star-Product}}
%EndExpansion
the Fedosov star-product is defined by means of its algorithm. The
properties are discussed as well as how to formulate the Klein-Gordon
equation in general in DQ.

%TCIMACRO{%
%\TeXButton{Codimension One}{\hyperref[s:Codimension One]{Section~\ref*{s:Codimension One}}} }%
%BeginExpansion
\hyperref[s:Codimension One]{Section~\ref*{s:Codimension One}}
%EndExpansion
states the original results of this paper. Beginning with the background
geometry and a phase-space connection we construct the Fedosov star for the
phase-space of any constant curvature manifold of codimension one. This
class of manifolds include the two-sphere, dS, and AdS. The background
geometry is reviewed as well as a phase-space connection introduced.

This section will read as follows: Each subsection (excluding the background
geometry subsection and the last three subsections of this section) will
remain completely general for an arbitrary phase-space until the
sub-subsection entitled: "The Constant Curvature Case Explicitly". It is
here we will state results specifically for the constant curvature manifold
case of codimension one. It is in the part of the subsection preceeding this
we will derive some general formulas and so will be valid for all
finite-dimensional phase-spaces.

\section{The Fedosov Star-Product}

%TCIMACRO{%
%\TeXButton{\label{s:The Fedosov Star-Product}}{\label{s:The Fedosov Star-Product}}}%
%BeginExpansion
\label{s:The Fedosov Star-Product}%
%EndExpansion
On a flat phase-space the Weyl quantization map $\mathcal{W}$ is the
isomorphism between the algebra of observables on a Hilbert space and the
Groenewold-Moyal star-algebra on phase-space. The goal of the Fedosov
algorithm is to construct a similar map called $\sigma ^{-1}$ on a general
phase-space which associates a unique Hilbert space operator $\hat{f}$ to
each phase-space function $f$. The map $\sigma ^{-1}$ in 
%TCIMACRO{%
%\TeXButton{Fedosov B. (1996)}{\hyperlink{ref1}{Fedosov B. (1996)}} }%
%BeginExpansion
\hyperlink{ref1}{Fedosov B. (1996)}
%EndExpansion
is a flat section in the Weyl-Heisenberg bundle (something which we will
define later). The star-product of any two phase-space functions would be
defined by:%
\begin{equation*}
f\ast g:=\sigma \left( \sigma ^{-1}\left( f\right) \sigma ^{-1}\left(
g\right) \right)
\end{equation*}%
analogously to the definition of the Groenewold-Moyal star-product $\left( 
\text{%
%TCIMACRO{\TeXButton{\ref{Moyal}}{\ref{Moyal}}}%
%BeginExpansion
\ref{Moyal}%
%EndExpansion
}\right) $. Fedosov provides an algorithm (see $\left[ \text{Fed}\right] $)
to construct the map $\sigma ^{-1}$ and $\sigma $. However, the construction
of such a map is a non-trivial task as we will see in the following
sections. With convergence issues aside, the properties of the Fedosov star
are (see 
%TCIMACRO{\TeXButton{Fedosov B. 1996}{\hyperlink{ref1}{Fedosov B. 1996}} }%
%BeginExpansion
\hyperlink{ref1}{Fedosov B. 1996}
%EndExpansion
and 
%TCIMACRO{%
%\TeXButton{Tillman P. and Sparling G. 2006}{\hyperlink{ref1}{Tillman P. and Sparling G. 2006}}}%
%BeginExpansion
\hyperlink{ref1}{Tillman P. and Sparling G. 2006}%
%EndExpansion
):

\begin{enumerate}
\item It is diffeomorphism covariant.

\item It can be constructed on all symplectic manifolds (including all
phase-spaces) perturbatively in powers of $\hbar $.

\item It assumes no dynamics (e.g. Hamiltonian or Lagrangian), symmetries,
or even a metric.

\item The limit $\hbar \rightarrow 0$ yields classical mechanics.

\item It is equivalent to an operator formalism by a Weyl-like quantization
map $\sigma ^{-1}$.
\end{enumerate}

In this paper we will restrict the focus onto phase-spaces of finite
dimensional manifolds because it these are the most relevant for the type of
physics we are interested in.

\bigskip

\noindent \textbf{Def.} A \textbf{symplectic manifold} is manifold equipped
with a non-degenerate (i.e., at all points $\omega _{AB}$ has an inverse $%
\omega ^{AB}$ st. $\omega ^{AB}\omega _{BC}=\delta _{C}^{A}$) closed
two-form.

It is well-known that all phase-spaces are symplectic manifolds. Consider $%
T^{\ast }%
%TCIMACRO{\U{211d} }%
%BeginExpansion
\mathbb{R}
%EndExpansion
^{n}$, the phase-space of $%
%TCIMACRO{\U{211d} }%
%BeginExpansion
\mathbb{R}
%EndExpansion
^{n}$. Choose the coordinates of the configuration space $%
%TCIMACRO{\U{211d} }%
%BeginExpansion
\mathbb{R}
%EndExpansion
^{n}$ to be $x^{\mu }$ then there exists canonical momentum associated to
these coordinates $p_{\mu }$. In these coordinates of phase-space $\left(
x,p\right) $ the symplectic form is $\omega =dp_{\mu }dx^{\mu
}=dp_{1}dx^{1}+\cdots +dp_{n}dx^{n}$. The Poisson bracket is then $\frac{%
\partial }{\partial x^{\mu }}\wedge \frac{\partial }{\partial p_{\mu }}$.

\bigskip

\noindent \textbf{Def.} The \textbf{cotangent space} $T_{x}^{\ast }M$ of a
manifold $M$ at the point $x\in M$ is the vector space of all possible
momenta $p_{\mu }$.

\bigskip

\noindent \textbf{Def.} The \textbf{cotangent bundle }or \textbf{phase-space 
}of $M$ is $T^{\ast }M=\cup _{x\in M}T_{x}^{\ast }M$ of a manifold $M$ is
the union of all tangent spaces at all points $x\in M$. A point in this
space is represented by $\left( x,p\right) $.

\bigskip

For this paper let $M$ be space-time. It is a fact for any $M$ that $T^{\ast
}M$ is always equipped with a nondegenerate closed two-form $\omega $ which
is basically the inverse of the Poisson bracket tensor.\footnote{%
The Poisson bracket tensor has two upstairs indices so is a $\left(
2,0\right) $ tensor and the symplectic form is a $\left( 0,2\right) $ tensor.%
} This is a straightforward generalization of the above example in $%
%TCIMACRO{\U{211d} }%
%BeginExpansion
\mathbb{R}
%EndExpansion
^{n}$ because we always have canonical momenta associated to each choice of
coordinates $x^{\mu }$. Therefore, every phase-space is a symplectic
manifold. The symplectic form in some local coordinates $\left( x,p\right) $
is $\omega =dp_{\mu }dx^{\mu }$ where $x$ is the coordinate on $M$ and $p$
is the canonical momentum conjugate to $x$. Also, on every phase-space we
can define a phase-space connection $D$ which we will need for the
construction of the Fedosov star-product. We define the Fedosov triple by $%
\left( T^{\ast }M,\omega ,D\right) $.

For any Fedosov triple Fedosov gives a perturbative expansion for a
generalized Groenewold-Moyal star-product we call the Fedosov star-product.
We note here that since the star-product is formulated in terms of a
perturbative expansion it's convergence issues remain unknown in general.

\subsection{The Klein-Gordon (KG) Equation on an Arbitrary Space-Time}

In order to gain a basic feel for this new formulation of quantum mechanics
we should re-express the fundamental quantities and equations into it. Here
we express the Klein-Gordon equation into this new language, i.e., into DQ.
In Minkowski space this is done by the use of the isomorphism of the Weyl
quantization map $\mathcal{W}$.

In special relativistic mechanics on Minkowski space the quantization of a
single particle begins with the classical invariant:%
\begin{equation*}
p_{\mu }p^{\mu }-m^{2}=0
\end{equation*}%
This invariant is then promoted to a constraint on the set of physically
allowed states where $m$ is the rest mass of the particle. The resulting
equation is the eigenvalue equation:%
\begin{equation*}
\left( \hat{p}_{\mu }\hat{p}^{\mu }-m^{2}\right) \left\vert \phi
_{m}\right\rangle =0~\ ~,~~~\left\langle \phi _{m}|\phi _{m}\right\rangle =1
\end{equation*}%
and computing:%
\begin{equation*}
\hat{H}\hat{\rho}_{m}=\hat{\rho}_{m}\hat{H}=m^{2}\hat{\rho}_{m}~~\
,~~~Tr\left( \hat{\rho}_{m}\right) =1~~~,~~~\hat{\rho}_{m}^{\dag }=\hat{\rho}%
_{m}~~\ ~,~~~~\hat{\rho}_{m}^{2}=\hat{\rho}_{m}
\end{equation*}%
where $\hat{H}=\hat{p}_{\mu }\hat{p}^{\mu }$, $\hat{\rho}_{m}:=\left\vert
\phi _{m}\right\rangle \left\langle \phi _{m}\right\vert $, $Tr$\ is the
full trace, and $\left\vert \phi _{m}\right\rangle $\ is a state of a spin
zero particle.

This equation can then be mapped to phase-space by $\mathcal{W}^{-1}$:%
\begin{equation}
H\ast \rho _{m}=\rho _{m}\ast H=m^{2}\rho _{m}~~\ ,~~~Tr_{\ast }\left( \rho
_{m}\right) =1~~~,~~~\bar{\rho}_{m}=\rho _{m}~~\ ~,~~~~\rho _{m}\ast \rho
_{m}=\rho _{m}
\end{equation}%
\begin{equation}
H=p_{\mu }\ast p^{\mu }
\end{equation}%
where $\ast $ is the Groenewold-Moyal star-product, $g_{\mu \nu }\left(
x\right) $ is the configuration space metric, $H=p_{\mu }p^{\mu }$ ($p^{\mu
}:=g^{\mu \nu }p_{\nu }$) and $\rho _{m}$ is the function that represents an
eigenstate of $H$.

In an analogous derivation (and by adding an arbitrary Ricci term\footnote{%
The reason we add an arbitrary Ricci term is because that we can't unallow
it. This term is standard in many texts like 
%TCIMACRO{%
%\TeXButton{Birrell N. and Davies P. 1982}{\hyperlink{ref1}{Birrell N. and Davies P. 1982}}}%
%BeginExpansion
\hyperlink{ref1}{Birrell N. and Davies P. 1982}%
%EndExpansion
.}) we can formulate the KG equation on an arbitrary space-time in DQ using
the map $\sigma ^{-1}$ provided by Fedosov's algorithm. $H$ is now replaced
with a new $H=p_{\mu }\ast p^{\mu }+\xi R$ where $R=R\left( x\right) $ is
the Ricci curvature scalar associated to this metric $g_{\mu \nu }\left(
x\right) $ of the space-time, $\xi \in 
%TCIMACRO{\U{2102} }%
%BeginExpansion
\mathbb{C}
%EndExpansion
$ is an arbitrary constant, and $\ast $ is now the Fedosov star-product.

The equation:%
\begin{equation}
\left( \hat{p}_{\mu }\hat{p}^{\mu }+\xi \hat{R}-m^{2}\right) \left\vert \phi
_{m}\right\rangle =0~\ ~,~~~\left\langle \phi _{m}|\phi _{m}\right\rangle =1
\label{KG}
\end{equation}%
becomes:%
\begin{equation}
H\ast \rho _{m}=\rho _{m}\ast H=m^{2}\rho _{m}~~\ ,~~~Tr_{\ast }\left( \rho
_{m}\right) =1~~~,~~~\bar{\rho}_{m}=\rho _{m}~~\ ~,~~~~\rho _{m}\ast \rho
_{m}=\rho _{m}  \label{KGDQ}
\end{equation}%
\begin{equation}
H=p_{\mu }\ast p^{\mu }+\xi R  \label{H}
\end{equation}

\subsection{The Algorithm}

%TCIMACRO{\TeXButton{\label{s:The Algorithm}}{\label{s:The Algorithm}}}%
%BeginExpansion
\label{s:The Algorithm}%
%EndExpansion
In this section we provide a brief outline of the algorithm that is used to
construct the Fedosov star-product. Because some of the formulas are put
into more convenient forms, constraints are carried through, as well as many
other complications, we want illustrate what the algorithm does.

\begin{description}
\item[Step 1.] 
%TCIMACRO{\TeXButton{\hypertarget{step 1}{}}{\hypertarget{step 1}{}}}%
%BeginExpansion
\hypertarget{step 1}{}%
%EndExpansion
We begin with a phase-space connection $D$:%
\begin{equation*}
Df=df=\frac{\partial f}{\partial x^{\mu }}dx^{\mu }+\frac{\partial f}{%
\partial p_{\mu }}dp_{\mu }
\end{equation*}%
\begin{equation*}
D\otimes \Theta ^{A}=\Gamma _{~B}^{A}\otimes \Theta ^{B}=\Gamma
_{~BC}^{A}\Theta ^{C}\otimes \Theta ^{B}
\end{equation*}%
where $\Theta ^{A}$ is a basis of one-forms in the cotangent bundle of our
phase-space (for example let $\Theta ^{A}=\left( dx^{\mu },dp_{\mu }\right) $%
). The symbol $\Gamma _{~BC}^{A}$ is defined to be the Christoffel symbol.
The connection preserves the symplectic two-form $\omega =\omega _{AB}\Theta
^{A}\wedge \Theta ^{B}$ (the inverse of the Poisson bracket tensor $\omega
^{AB}$, i.e., $\omega ^{AB}\omega _{BC}=\delta _{C}^{A}$) by $D\otimes
\omega =0$. In the coordinates $\left( x^{\mu },p_{\mu }\right) $ $\omega
=dp_{\mu }\wedge dx^{\mu }$. The Poisson bracket operator is $\omega ^{AB}%
\frac{\partial }{\partial q^{A}}\wedge \frac{\partial }{\partial q^{B}}$.

\item[Step 2.] 
%TCIMACRO{\TeXButton{\hypertarget{step 2}{}}{\hypertarget{step 2}{}}}%
%BeginExpansion
\hypertarget{step 2}{}%
%EndExpansion
To each point $q=\left( x,p\right) $ on the phase-space we associate a
matrix algebra called the Heisenberg-Weyl algebra. The union of these
algebras is called the Weyl-Heisenberg bundle over the phase-space. We
define the basis elements $\hat{y}^{A}$ as an infinite-dimensional matrix. $%
\hat{y}^{A}$ is defined to have the properties:%
\begin{equation}
\left[ \hat{y}^{A},\hat{y}^{B}\right] =\hat{y}^{B}\hat{y}^{A}-\hat{y}^{B}%
\hat{y}^{A}=i\hbar \omega ^{AB}\hat{1}  \label{yy}
\end{equation}%
\begin{equation}
D\hat{y}^{A}=\Gamma _{~BC}^{A}\Theta ^{C}\hat{y}^{B}  \label{Dy}
\end{equation}%
where $\hat{1}$ is the identity matrix and it is assumed that $\Theta $ are
treated as a scalar with respect to $\hat{y}$'s matrix indices ($[\Theta
^{A},\hat{y}^{B}]=0$).

*Note that we will omit the $\hat{1}$ from the formula from now on and it is
implicitly there.

To better understand these $\hat{y}^{A}$ we should think of them as a matrix
with matrix-elements which are functions. Explicitly we have:%
\begin{equation*}
\hat{y}^{A}=\left( 
\begin{array}{ccc}
y_{11}^{A}\left( x,p\right) & y_{12}^{A}\left( x,p\right) & \cdots \\ 
y_{21}^{A}\left( x,p\right) & y_{22}^{A}\left( x,p\right) & \cdots \\ 
\vdots & \vdots & \ddots%
\end{array}%
\right)
\end{equation*}%
so that $y_{ij}^{A}\left( x,p\right) $ is a function for each $i$ and $j$.

\item[Step 3.] 
%TCIMACRO{\TeXButton{\hypertarget{step 3}{}}{\hypertarget{step 3}{}}}%
%BeginExpansion
\hypertarget{step 3}{}%
%EndExpansion
We define a matrix operator called $\hat{D}$ defined by the graded commutator%
\footnote{%
Graded commutators have the property that $\left[ \hat{Q}_{A}\Theta ^{A},w%
\right] =\left[ \hat{Q}_{A},w\right] \Theta ^{A}=\left( \hat{Q}_{A}w-w\hat{Q}%
_{A}\right) \Theta ^{A}$ where $w$ is an arbitrary $l$-form with
coefficients $w_{A_{1}\cdots A_{l}}$ which are complex-valued functions of
the variables $x,p$ and $\hat{y}$.}:%
\begin{equation}
\hat{D}=\left[ \hat{Q},\cdot \right] /i\hbar =\left[ \hat{Q}_{A}\Theta
^{A},\cdot \right] /i\hbar  \notag
\end{equation}%
\begin{equation*}
\hat{Q}_{A}=\sum_{l}Q_{AA_{1}\cdots A_{l}}\hat{y}^{A_{1}}\cdots \hat{y}%
^{A_{l}}
\end{equation*}%
where $Q_{AA_{1}\cdots A_{l}}$\ are complex-valued functions of $x$ and $p$
that need to be determined.

The coefficients $Q_{AA_{1}\cdots A_{l}}$ are partially determined\footnote{%
Fedosov adds an additional condition that makes his $\hat{D}$ unique from a
fixed $D$ being $\hat{d}^{-1}r_{0}=0$ where $\hat{d}^{-1}$ is what he calls $%
\delta ^{-1}$ (an operator used in a de Rham decomposition) and $r_{0}$ is
the first term in the recursive solution. We regard this choice as being
artificial and thus omit it from the paper.} by the condition:%
\begin{equation}
\left( D-\hat{D}\right) ^{2}\hat{y}^{A}=0  \label{cond_Dhat}
\end{equation}%
We can fix $\hat{D}$ any way we like, just as long as the above condition
holds. The way to think of the above condition is as an integrability
condition in the construction of the observable algebra.

\item[Step 4.] 
%TCIMACRO{\TeXButton{\hypertarget{step 4}{}}{\hypertarget{step 4}{}}}%
%BeginExpansion
\hypertarget{step 4}{}%
%EndExpansion
We then use $\hat{D}$ to define the algebra of observables to be the set of
all functions $\hat{f}$:%
\begin{equation}
\hat{f}\left( x,p,\hat{y}\right) =\sum_{j,l}f_{j,l,A_{1}\cdots A_{l}}\hbar
^{j}\hat{y}^{A_{1}}\cdots \hat{y}^{A_{l}}  \label{fhat}
\end{equation}%
where $f_{j,l,A_{1}\cdots A_{l}}$\ are complex-valued functions of $x$ and $%
p $ \ for each $j,l,A_{1},\ldots ,A_{l}$ that need to be determined.
Moreover the indices $\left( A_{1}\cdots A_{l}\right) $ \ \ are assumed to
be symmetric.

For every function $f\left( x,p\right) $ the coefficients $%
f_{j,l,A_{1}\cdots A_{l}}$ \ of the series above are partially determined by
the conditions:%
\begin{equation}
\left( D-\hat{D}\right) \hat{f}=0  \label{cond_fhat}
\end{equation}%
\begin{equation*}
\sigma (\hat{f})=f_{0,0}=f\left( x,p\right)
\end{equation*}%
where $\sigma $ is defined to be:%
\begin{equation*}
\sigma (\hat{f})=\sum_{j,l}f_{j,0}\hbar ^{j}
\end{equation*}%
We can make any choice that fixes the additional freedom and in total this
gives us the map we need $\sigma ^{-1}$.

\noindent \textbf{Note:} The inverse quantization map $\sigma $ is defined
to be the operation that picks out the leading order term in the symmetrized
series for $f$ in $\left( \text{%
%TCIMACRO{\TeXButton{\ref{fhat}}{\ref{fhat}}}%
%BeginExpansion
\ref{fhat}%
%EndExpansion
}\right) $, i.e., the term that has no $\hat{y}$'s in them.

\item[Step 5.] 
%TCIMACRO{\TeXButton{\hypertarget{step 5}{}}{\hypertarget{step 5}{}}}%
%BeginExpansion
\hypertarget{step 5}{}%
%EndExpansion
The Fedosov star-product $f\ast g$ is defined by:%
\begin{equation*}
f\ast g:=\sigma (\sigma ^{-1}\left( f\right) \sigma ^{-1}\left( g\right)
)=\sigma (\hat{f}\hat{g})
\end{equation*}%
*Note that to get the leading order term $\sigma (\hat{f}\hat{g})$ you have
to symmetrize all the monomials in $\hat{y}$'s in the product $\hat{f}\hat{g}
$ first, then take the leading term. This makes the multiplication of $f\ast
g$ highly non-trivial.
\end{description}

There is some freedom in choosing $D$ and $\hat{D}$ but once they are chosen
we can associate unique operators $\hat{f}$ to every phase-space function.
This is precisely the map that we need $\sigma ^{-1}$ (and $\sigma $), $%
\sigma ^{-1}\left( f\right) =\hat{f}$ (i.e., $\sigma ^{-1}$ is a section in
the bundle). Moreover, the reason we call $\sigma ^{-1}$ a flat section
because it is constructed with the condition that the curvature of the
derivation $(D-\hat{D})$ is zero, the condition $\left( \text{%
%TCIMACRO{\TeXButton{\ref{cond_Dhat}}{\ref{cond_Dhat}}}%
%BeginExpansion
\ref{cond_Dhat}%
%EndExpansion
}\right) $.

\section{The Fedosov Star-Product on Constant Curvature Manifolds of
Codimension One}

%TCIMACRO{\TeXButton{\label{s:Codimension One}}{\label{s:Codimension One}}}%
%BeginExpansion
\label{s:Codimension One}%
%EndExpansion
Now that we are familiar with the basics of DQ and the Fedosov star-product,
we shall explicate the results of the paper. The focus of this paper is on a
particular star-product known as the Fedosov star-product. Fedosov
star-product is a star-product that can be written down at least in a formal
power series in $\hbar $ for any generalization of a phase-space of
arbitrary space-time manifold called a symplectic manifold. Although
symplectic manifolds are more general manifolds than phase-spaces, we will
only consider phase-spaces.

As stated before, the primary aim of the paper was to construct the Fedosov
star-product on the phase-space of a single particle in the case of all
(finite-dimensional) constant curvature manifolds embeddable in a flat space
with codimension 1. The observable algebra is algebra of functions on
phase-space along with this new product. This set of spaces includes the
two-sphere and de Sitter (dS)/anti-de Sitter (AdS) space-times. By
techniques provided by Fedosov's algorithm we can construct the quantization
map $\sigma ^{-1}$ (and also $\sigma $ but this map is trivial to construct
so all of our hard work goes into $\sigma ^{-1}$) which is what these
results do. The crucial ingredient is the construction of a new dervation $%
\hat{D}$ so that $\left( D-\hat{D}\right) $ is a flat derivation in 
%TCIMACRO{%
%\TeXButton{\hyperlink{step 3}{\textbf{step 3}}}{\hyperlink{step 3}{\textbf{step 3}}}}%
%BeginExpansion
\hyperlink{step 3}{\textbf{step 3}}%
%EndExpansion
. This derivation is crucial to the definition of the algebra in 
%TCIMACRO{%
%\TeXButton{\hyperlink{step 4}{\textbf{step 4}}}{\hyperlink{step 4}{\textbf{step 4}}}}%
%BeginExpansion
\hyperlink{step 4}{\textbf{step 4}}%
%EndExpansion
. From this we can write down the star-product for any phase-space functions
in powers of $\hbar $.

The purpose of these results was four-fold. One was to verify that DQ gave
the same results as previous analyses of these spaces. Another was to verify
that the formal series obtained by the Fedosov algorithm converged by
obtaining \textit{exact} and \textit{nonperturbative} results for these
spaces. As was stated in the introduction, one the most serious issues
confronting DQ is the issue of convergence of all formal series in $\hbar $.
Therefore, if the star-product has any merit at all in describing quantum
theories on non-trivial manifolds it should be well-defined for some of the
simplest cases, i.e., constant curvature manifolds.

The last goal was to further develop the technology of the Fedosov
algorithm. This includes developing a refinement of the formulas for the
algorithm by assuming that the symplectic manifold is a phase-space. We then
show that the resulting condition $\left( \text{%
%TCIMACRO{\TeXButton{\ref{cond Qhat}}{\ref{cond Qhat}}}%
%BeginExpansion
\ref{cond Qhat}%
%EndExpansion
}\right) $ is locally integrable by the Cauchy-Kovalevskaya theorem.

This section will read as follows: Each subsection (excluding the background
geometry subsection and the last three subsections of this section) will
remain completely general for an arbitrary phase-space until the
sub-subsection entitled: "The Constant Curvature Case Explicitly". It is
here we will state results specifically for the constant curvature manifold
case of codimension one. It is in the part of the subsection preceeding this
we will derive some general formulas and so will be valid for all
finite-dimensional phase-spaces.

\subsection{The Background Geometry}

Before we go into the details of the results we first want to review the
geometry of constant curvature manifolds of codimension one. To this end, we
rely on the fact that it is a relatively straightforward generalization of
the familiar two-sphere and dS/AdS manifolds. The fact that the sphere and
dS/AdS lie in this class is the main motivation for considering it.

We start with the phase space of a single classical particle confined to a
constant curvature manifold with metric $\left( M_{C_{p,q}},g\right) $ that
is imbedded in $\left( 
%TCIMACRO{\U{211d} }%
%BeginExpansion
\mathbb{R}
%EndExpansion
^{n+1},\eta \right) $ where $\dim M_{C_{p,q}}=p+q=n$ and $\eta $ is a
pseudoeuclidean metric. The imbedding specifically is the hyperboloid:%
\begin{equation*}
x^{\mu }x_{\mu }=\eta _{\mu \nu }x^{\mu }x^{\nu }=1/C
\end{equation*}%
$\eta $ induces a metric on $M_{C_{p,q}}$ called$\ g$\ and explicitly:%
\begin{equation}
g_{\mu \nu }:=\eta _{\mu \nu }-Cx_{\mu }x_{\nu }  \label{g0}
\end{equation}%
which is easily obtained by the constraint above (just project each index
orthogonal to $x$). Also, we will always raise and lower the lower-case
indices or $M_{C_{p,q}}$\ indices (greek or latin) by the metric of the
imbedding space $%
%TCIMACRO{\U{211d} }%
%BeginExpansion
\mathbb{R}
%EndExpansion
^{n+1}$ $\eta $.

We make the convention that the positive signature directions are the "time"
directions while the negative ones are the "space" directions. If the
signature of $g$ denoted by $sign\left( g\right) $ is $\left( p,q\right) $\
then for $C>0$ (this space-time is denoted by $M_{C_{p,q}}^{+}$), $\eta $ is
a pseudoeuclidean metric of signature $\left( p+1,q\right) $ or explicitly:%
\begin{equation*}
\eta =diag\underset{p+1}{(\underbrace{1,\ldots ,1}},\underset{q}{\underbrace{%
-1,\ldots ,-1})}
\end{equation*}%
If, however, $C<0$ (this space-time is denoted by $M_{C_{p,q}}^{-}$), $\eta $
is a pseudo-euclidean metric of signature $\left( p,q+1\right) $. This is
because for $C>0$ the hyperboloid is "time"-like, i.e., it has normal
vectors pointing in a combination of the $p+1$ positive signature directions
thus the induced metric has a signature of one less "time" dimensions from
the imbedding. For the case of $C<0$ the hyperboloid is space-like and thus
the induced metric has a signature of one less "space" dimensions, i.e., it
has normal vectors pointing in a combination of the $q+1$ negative signature
directions.

A good way to visualize these spaces is to look at the $1+3$ dimensions
which gives us the familiar de Sitter (dS) and Anti-de Sitter (AdS)
space-times for $C<0$ and $C>0$ respectively. The picture, of course,
generalizes very naturally. The embeddings in these cases are:%
\begin{equation}
\left( x^{0}\right) ^{2}-\left( x^{4}\right) ^{2}-\underline{x}\cdot 
\underline{x}=1/C~~~,~~C<0
\end{equation}%
\begin{equation}
\left( x^{0}\right) ^{2}+\left( x^{4}\right) ^{2}-\underline{x}\cdot 
\underline{x}=1/C~~,~~C>0
\end{equation}%
where:%
\begin{equation*}
\underline{x}=\left( x^{1},x^{2},x^{3}\right) 
\end{equation*}%
We notice that in the case of dS the definition of time must be $x^{0}$ and
in AdS it must be the 0-4 angle $\theta $. We immediately notice a problem
in this embedding of AdS: If we follow a world line starting at $\theta =0$
and ending at $\theta =2\pi $ we arrive back at our starting point. We
reason that we cannot reach the past by going far into the future. This is
to avoid serious paradoxes of what must be a pathological space-time.

The resolution to this dilemma is to go to the covering space of the
hyperboloid by "unidentifying" (or not identifying them in the first place)
the values $0$, $\pm 2\pi $, $\pm 4\pi $,$\ldots $. This is done by breaking
the hyperboloid into leaves (labelled by $n$) and so if we follow a
world-line starting at $\theta =0$ when we get to $2\pi $ we\ will be in a
different leaf of the covering space and thus not at our original point. The
picture is described by first imagining that we have infinitely many
hyperboloids. We then cut them length-wise, open them up, and put each
successive one above the other. Thus the topology of time is $\mathbb{%
%TCIMACRO{\U{211d} }%
%BeginExpansion
\mathbb{R}
%EndExpansion
}$ not an $\mathbb{S}^{1}$.

By differentiating $x^{\mu }x_{\mu }=1/C$ we may obtain the condition on $%
p_{\mu }$:%
\begin{equation*}
2dx^{\mu }x_{\mu }=0\implies x^{\mu }p_{\mu }=0
\end{equation*}%
The embedding formulas are then: 
\begin{equation}
x^{\mu }x_{\mu }=1/C~~\ ,~~\ x^{\mu }p_{\mu }=0  \label{embedding}
\end{equation}%
where $C$ is an arbitrary real constant.

\subsection{The Phase-Space Connection}

The starting place of the Fedosov algorithm is the phase-space connection $D$
in 
%TCIMACRO{%
%\TeXButton{\hyperlink{step 1}{\textbf{step 1}}}{\hyperlink{step 1}{\textbf{step 1}}} }%
%BeginExpansion
\hyperlink{step 1}{\textbf{step 1}}
%EndExpansion
of the algorithm. In this section we construct a connection suitable for our
purposes although any could be chosen. We choose a torsion-free phase-space
connection that preserves the metric. To construct $D$ we start with the
Levi-Civita connection $\nabla $ on the configuration space $M$ and use this
to derive the desired phase-space connection.

We now introduce a Levi-Civita connection $\nabla $ on the configuration
space $M$ and subsequent curvature given the metric $g$ on a general
manifold $M$:%
\begin{equation}
\nabla _{\sigma }f\left( x\right) =\frac{\partial f}{\partial x^{\sigma }}
\label{nabla}
\end{equation}%
\begin{equation*}
\nabla _{\sigma }\left( dx^{\mu }\right) =-\Gamma _{~\nu \sigma }^{\mu
}dx^{\nu }
\end{equation*}%
\begin{equation*}
\nabla _{\sigma }\left( \frac{\partial }{\partial x^{\mu }}\right) =\Gamma
_{~\mu \sigma }^{\nu }\frac{\partial }{\partial x^{\nu }}
\end{equation*}%
\begin{equation*}
\nabla _{\lbrack \sigma }\nabla _{\rho ]}\left( dx^{\mu }\right) =R_{~\nu
\sigma \rho }^{\mu }dx^{\nu }
\end{equation*}%
where $R_{~\nu \sigma \rho }^{\mu }$ is the Riemann tensor. Of course we
have the conditions that $\nabla $ preserves the metric $g$ and is
torsion-free:%
\begin{equation*}
\nabla _{a}g_{bc}=0
\end{equation*}%
\begin{equation*}
\nabla _{\lbrack a}\nabla _{b]}f\left( x\right) =0
\end{equation*}%
for all functions $f\left( x\right) $. Together these uniquely fix $\nabla $.

We can "lift" the action of this connection $\nabla $ to induce a unique
phase-space connection $D$ (see 
%TCIMACRO{%
%\TeXButton{\hyperlink{Appendix D}{appendix D}}{\hyperlink{Appendix D}{appendix D}} }%
%BeginExpansion
\hyperlink{Appendix D}{appendix D}
%EndExpansion
for the details). The way we can think of this induction is that the
configuration space connection $\nabla $ acts naturally on the covectors of
covectors (which are essentially two-index tensors). On the cotangent bundle
of phase-space we define a basis of one-forms $\left( dx^{\mu },\alpha _{\mu
}\right) $ where $\alpha _{\mu }$ is defined as:

\begin{equation}
\alpha _{\mu }:=dp_{\mu }-\Gamma _{~\mu \rho }^{\nu }dx^{\rho }p_{\nu }
\label{alpha}
\end{equation}%
\pagebreak 

\noindent We define the phase-space connection to be:%
\begin{equation}
Dx^{\mu }:=dx^{\mu }  \label{D}
\end{equation}%
\begin{equation*}
Dp_{\mu }:=dp_{\mu }
\end{equation*}%
\begin{equation*}
D\otimes dx^{\mu }=-\Gamma _{~\sigma \nu }^{\mu }dx^{\nu }\otimes dx^{\sigma
}
\end{equation*}%
\begin{equation*}
D\otimes \alpha _{\mu }=\Theta ^{B}\otimes D_{B}\alpha _{\mu }:=-\frac{4}{3}%
R_{~(\mu \sigma )\beta }^{\psi }p_{\psi }dx^{\beta }\otimes dx^{\sigma
}+\Gamma _{~\mu \sigma }^{\nu }dx^{\sigma }\otimes \alpha _{\nu }
\end{equation*}%
\begin{equation*}
\alpha _{\mu }:=dp_{\mu }-\Gamma _{~\mu \rho }^{\nu }dx^{\rho }p_{\nu }
\end{equation*}%
and the corresponding curvature:%
\begin{equation}
D^{2}x^{\mu }=0  \label{D2}
\end{equation}%
\begin{equation*}
D^{2}p_{\mu }=0
\end{equation*}%
\begin{equation}
D^{2}\otimes dx^{\mu }=dx^{\sigma }dx^{\rho }\otimes R_{~\nu \sigma \rho
}^{\mu }dx^{\nu }  \notag
\end{equation}%
\begin{equation}
D^{2}\otimes \alpha _{\mu }=\frac{4}{3}dx^{\sigma }\left( C_{~\mu \beta \nu
\sigma }^{\psi }p_{\psi }dx^{\nu }+R_{~(\mu \beta )\sigma }^{\nu }\alpha
_{\nu }\right) \otimes dx^{\beta }-R_{~\mu \sigma \beta }^{\nu }dx^{\sigma
}dx^{\beta }\otimes \alpha _{\nu }  \notag
\end{equation}%
where $C_{~abes}^{c}:=\nabla _{s}R_{~(ab)e}^{c}$ and according to $\left( 
\text{%
%TCIMACRO{\TeXButton{\ref{nabla}}{\ref{nabla}}}%
%BeginExpansion
\ref{nabla}%
%EndExpansion
}\right) $ the formula for the curvature is:%
\begin{equation}
R_{~\nu \sigma \rho }^{\mu }=-\partial _{\lbrack \sigma }\Gamma _{~\rho ]\nu
}^{\mu }+\Gamma _{~\nu \lbrack \sigma }^{\kappa }\Gamma _{~\rho ]\kappa
}^{\mu }  \label{Curvature}
\end{equation}%
We can extend to higher order tensors by using the Leibnitz rule and the
fact that $D$ and $\nabla $ commute with contractions.

\subsubsection{The Constant Curvature Case Explicitly}

Given a configuration space connection $\nabla $ it was a relatively
straight forward matter to derive a phase-space connection associated to it.
So all we need formulas for the Christhoffel symbols $\Gamma $ in our
coordinates and we're done. Normally this would be a straightforward matter,
but because of the constraints:%
\begin{equation}
x^{\mu }x_{\mu }=1/C~~~~,~~~x^{\mu }p_{\mu }=0  \label{constraintA}
\end{equation}%
and the subsequent conditions:%
\begin{equation}
x^{\mu }dx_{\mu }=0~~~~,~~~p_{\mu }dx^{\mu }+x^{\mu }dp_{\mu }=0
\label{constraintB}
\end{equation}%
the situation becomes a bit more muddled.

Without constraints when given a metric the Levi-Civita (torsion-free and
metric preserving) and its curvature would be determined uniquely by the
formulas $\left( \text{%
%TCIMACRO{\TeXButton{\ref{Gamma}}{\ref{Gamma}}}%
%BeginExpansion
\ref{Gamma}%
%EndExpansion
}\right) $ and $\left( \text{%
%TCIMACRO{\TeXButton{\ref{Curvature}}{\ref{Curvature}}}%
%BeginExpansion
\ref{Curvature}%
%EndExpansion
}\right) $ in 
%TCIMACRO{%
%\TeXButton{\hyperlink{Appendix D}{appendix D}}{\hyperlink{Appendix D}{appendix D}}}%
%BeginExpansion
\hyperlink{Appendix D}{appendix D}%
%EndExpansion
. However, when we compute them using these formulas we are still left with
freedom resulting from the above constraint equations. A particular formula
like:%
\begin{equation*}
D\otimes dx^{\mu }=-\Gamma _{~\sigma \nu }^{\mu }dx^{\nu }\otimes dx^{\sigma
}
\end{equation*}%
is obviously ambiguous because under the constraint $x^{\mu }dx_{\mu }=0$ in 
$\left( \text{%
%TCIMACRO{\TeXButton{\ref{constraintB}}{\ref{constraintB}}}%
%BeginExpansion
\ref{constraintB}%
%EndExpansion
}\right) $ so that the formula above is invariant under the change:%
\begin{equation*}
\Gamma _{~\mu \nu }^{\rho }\rightarrow \Gamma _{~\mu \nu }^{\rho }+x^{\rho
}q_{\mu \nu }+x_{(\mu }f_{~\nu )}^{\rho }
\end{equation*}%
where $q_{\mu \nu }$ and $f_{~\nu }^{\rho }$ are arbitrary (the
symmetrization of $x_{(\mu }f_{~\nu )}^{\rho }$ is to preserve the
torsion-free condition).

The reason there is some freedom is because we need to additionally impose
that the connection preserves the above conditions.\footnote{%
To be technically correct, the constraints $\left( \text{%
%TCIMACRO{\TeXButton{\ref{constraintB}}{\ref{constraintB}}}%
%BeginExpansion
\ref{constraintB}%
%EndExpansion
}\right) $ come from the connection's action on the constraints $\left( 
\text{%
%TCIMACRO{\TeXButton{\ref{constraintA}}{\ref{constraintA}}}%
%BeginExpansion
\ref{constraintA}%
%EndExpansion
}\right) $.} This will subsequently fix most of the additional
freedom.\pagebreak 

We then require that these constraints are preserved by the connection:%
\begin{equation}
D\left( x^{\mu }x_{\mu }\right) =0~~~,~~~D\left( x^{\mu }p_{\mu }\right)
=0~~~,~~~D^{2}\left( x^{\mu }x_{\mu }\right) =0~~~,~~~D^{2}\left( x^{\mu
}p_{\mu }\right) =0  \label{Dconstraint}
\end{equation}%
as well as equations coming from higher order derivatives.

The way will proceed is first compute the connection and curvature using $%
\left( \text{%
%TCIMACRO{\TeXButton{\ref{Gamma}}{\ref{Gamma}}}%
%BeginExpansion
\ref{Gamma}%
%EndExpansion
}\right) $, $\left( \text{%
%TCIMACRO{\TeXButton{\ref{Curvature}}{\ref{Curvature}}}%
%BeginExpansion
\ref{Curvature}%
%EndExpansion
}\right) $, the ambient connection $\partial $ and the formula for the
metric in $\left( \text{%
%TCIMACRO{\TeXButton{\ref{g0}}{\ref{g0}}}%
%BeginExpansion
\ref{g0}%
%EndExpansion
}\right) $. We then fix the additional freedom by imposing the constraints
in $\left( \text{%
%TCIMACRO{\TeXButton{\ref{Dconstraint}}{\ref{Dconstraint}}}%
%BeginExpansion
\ref{Dconstraint}%
%EndExpansion
}\right) $. We will be left with a little additional freedom which will not
affect any of our formulas so we make an arbitrary choice here. The result
will give us the formulas in $\left( \text{%
%TCIMACRO{\TeXButton{\ref{g}}{\ref{g}}}%
%BeginExpansion
\ref{g}%
%EndExpansion
}\right) $.

The conditions that $\Gamma $ must satisfy are:

\begin{enumerate}
\item torsion-free:%
\begin{equation*}
dx^{\sigma }\nabla _{\sigma }\left( dx^{\mu }\right) =-\Gamma _{~\nu \sigma
}^{\mu }dx^{\sigma }dx^{\nu }\implies \Gamma _{~[\nu \sigma ]}^{\mu }=0
\end{equation*}

\item metric-preserving:%
\begin{equation*}
\nabla _{\rho }\left( g_{\mu \nu }dx^{\mu }dx^{\nu }\right) =0
\end{equation*}

\item The directional derivative $\mathcal{D}_{v}$ of a vector and covector
in any direction $v^{a}$ is also a vector and covector respectively.%
\begin{equation*}
w_{\mu }\text{ is a covector }\iff \mathcal{D}_{v}w_{\mu }=v^{\rho }\left(
\partial _{\rho }w_{\mu }-\Gamma _{~\mu \rho }^{\nu }w_{\nu }\right) \text{
is a covector}
\end{equation*}%
\begin{equation*}
w^{\mu }\text{ is a vector }\iff \mathcal{D}_{v}w^{\mu }=v^{\rho }\left(
\partial _{\rho }w^{\mu }+\Gamma _{~\nu \rho }^{\mu }w^{\nu }\right) \text{
is a vector}
\end{equation*}

\item The constraints in $\left( \text{%
%TCIMACRO{\TeXButton{\ref{constraintA}}{\ref{constraintA}}}%
%BeginExpansion
\ref{constraintA}%
%EndExpansion
}\right) $ and $\left( \text{%
%TCIMACRO{\TeXButton{\ref{constraintB}}{\ref{constraintB}}}%
%BeginExpansion
\ref{constraintB}%
%EndExpansion
}\right) $:%
\begin{equation}
x^{\mu }x_{\mu }=1/C~~~,~~~x^{\mu }p_{\mu }=0  \label{constraintA0}
\end{equation}%
\begin{equation}
x_{\mu }dx^{\mu }=0~~~,~~~~dx^{\mu }p_{\mu }+x^{\mu }dp_{\mu }=0
\label{constraintB0}
\end{equation}%
\begin{equation*}
\nabla _{\nu }\left( x^{\mu }dx_{\mu }\right) =0~~~,~~~\nabla _{\nu }\left(
p_{\mu }dx^{\mu }+x^{\mu }dp_{\mu }\right) =0
\end{equation*}
\end{enumerate}

The configuration space metric, Christhoffel symbol and Riemann tensor are
for our specific case $M_{C_{p,q}}$ (and for our choice of coordinates) are
using the above strategy: 
\begin{equation}
g_{\mu \nu }=\eta _{\mu \nu }-Cx_{\mu }x_{\nu }  \label{g}
\end{equation}%
\begin{equation*}
\Gamma _{~\nu \sigma }^{\mu }=Cx^{\mu }g_{\nu \sigma }-2Cx_{(\nu }\left(
\delta _{\sigma )}^{\mu }-Cx_{\sigma )}x^{\mu }\right)
\end{equation*}%
\begin{equation}
R_{~\nu \sigma \rho }^{\mu }=-C\left( \delta _{\lbrack \sigma }^{\mu
}-Cx_{[\sigma }x^{\mu }\right) g_{\rho ]\nu }  \notag
\end{equation}%
\begin{equation*}
\omega =\left( \delta _{\nu }^{\mu }-Cx^{\mu }x_{\nu }\right) \alpha _{\mu
}dx^{\nu }
\end{equation*}%
On a general techinical note, we will proceed in an identical fashion for
most of the paper: a each step verify that all relelvant constraints are
satisfied. Although we choose a set of coordinates, even ones with
constraints $x^{\mu }$, the objects we consider such as $\nabla $, $g$, etc.
are intrinsic and coordinate independent things.

\subsection{The Weyl-Heisenberg Bundle}

In 
%TCIMACRO{%
%\TeXButton{\hyperlink{step 2}{\textbf{step 2}}}{\hyperlink{step 2}{\textbf{step 2}}} }%
%BeginExpansion
\hyperlink{step 2}{\textbf{step 2}}
%EndExpansion
of the algorithm, we introduce some machinery namely the operators $\hat{y}$%
's to calculate the observables on general manifold $M$. However, unlike
Fedosov who defines these $\hat{y}$'s as covectors equipped with a
Moyal-like product between them we let these $\hat{y}$'s to be infinite
dimensional matrix-valued operators acting on a Hilbert space. The relations
defining the $\hat{y}$'s (relations $\left( \text{%
%TCIMACRO{\TeXButton{\ref{yy}}{\ref{yy}}}%
%BeginExpansion
\ref{yy}%
%EndExpansion
}\right) $ and $\left( \text{%
%TCIMACRO{\TeXButton{\ref{Dy}}{\ref{Dy}}}%
%BeginExpansion
\ref{Dy}%
%EndExpansion
}\right) $ in 
%TCIMACRO{%
%\TeXButton{\hyperlink{step 2}{\textbf{step 2}}}{\hyperlink{step 2}{\textbf{step 2}}}}%
%BeginExpansion
\hyperlink{step 2}{\textbf{step 2}}%
%EndExpansion
) are identical in both cases.

\noindent \underline{The Link to Familiar Heisenberg Algebras Using Darboux
Coordinates:}

The first relation $\left( \text{%
%TCIMACRO{\TeXButton{\ref{yy}}{\ref{yy}}}%
%BeginExpansion
\ref{yy}%
%EndExpansion
}\right) $ in 
%TCIMACRO{%
%\TeXButton{\hyperlink{step 2}{\textbf{step 2}}}{\hyperlink{step 2}{\textbf{step 2}}} }%
%BeginExpansion
\hyperlink{step 2}{\textbf{step 2}}
%EndExpansion
is $\left[ \hat{y}^{A},\hat{y}^{B}\right] =i\hbar \omega ^{AB}$ and can be
expressed in a more familiar form by a suitable choice of coordinates. In
symplectic geometry there is a famous theorem, called Darboux's theorem,
which states that in the neighborhood of each point on an $n$-dimensional
symplectic manifold, there exists coordinates called Darboux coordinates $%
\tilde{q}=\left( \tilde{x}^{1},\ldots ,\tilde{x}^{n},\tilde{p}_{1},\ldots ,%
\tilde{p}_{n}\right) $\footnote{%
Note that these $2n$ coordinates and are different from the $2n+2$ embedding
coordinates $\left( x^{\mu },p_{\mu }\right) $.} where the $\omega $ takes
the form:%
\begin{equation*}
\omega =d\tilde{p}_{1}d\tilde{x}^{1}+\cdots +d\tilde{p}_{n}d\tilde{x}^{n}
\end{equation*}%
In this coordinate system at $\tilde{q}$ the $\hat{y}$'s are expressed as $%
2n $\ operators $\left( \tilde{s}^{1},\ldots ,\tilde{s}^{n},\tilde{k}%
_{1},\ldots ,\tilde{k}_{n}\right) $ which have the commutators $\left[ 
\tilde{s}^{i},\tilde{s}^{j}\right] =\left[ \tilde{k}_{i},\tilde{k}_{j}\right]
=0,~\left[ \tilde{s}^{i},\tilde{k}_{j}\right] =i\hbar \delta _{j}^{i}$ where 
$i$ and $j$ run from $1$ through $2n$. And so at each point the $\hat{y}$'s
establish a standard Heisenberg algebra (acting on a Hilbert space) which
all physicists know. Therefore at each point we have a standard algebra of
observables that we are intimately familiar with in ordinary quantum
mechanics. The full bundle of all of these algebras at all points creates a
huge algebra and it is the goal of the Fedosov algorithm is to choose an
appropriate subalgebra in this huge algebra that we can identify as our
algebra of observables subject, of course, to agreement to real physical
situations. This subalgebra is the image of the map $\sigma ^{-1}$ on the
set of all phase-space functions.

\noindent \underline{Defining Properties of $\hat{y}$:}%
\begin{equation*}
\left[ \hat{y}^{A},\hat{y}^{B}\right] =i\hbar \omega ^{AB}
\end{equation*}%
\begin{equation*}
D\hat{y}^{A}=-\Gamma _{~B}^{A}\hat{y}^{B}=-\Gamma _{~BC}^{A}\Theta ^{C}\hat{y%
}^{B}~~~,~~~\Theta ^{B}=\left( \theta ^{\sigma },\alpha _{\sigma }\right)
\end{equation*}%
The $\hat{y}$'s commute with the set of quantities $\left\{ x,p,\Theta
,g,\omega ,\hbar ,i\right\} $ (i.e., they behave as scalars on the matrix
indices) where $i$ is the complex unit.

*Note that the action of the phase-space connection on $\hat{y}$ is the same
as the one on $\Theta $ ($D\otimes \Theta ^{A}=\Gamma _{~BC}^{A}\Theta
^{C}\otimes \Theta ^{B}$) and so we regard it as a basis of operator or
matrix-valued covectors. The connection's action on the $\hat{y}$'s tells us
how to parallel transport the Weyl-Heisenberg algebra (the $\hat{y}$'s) at
one point to the Weyl-Heisenberg algebra of every other point in a
consistent way.

By defining $\hat{y}^{A}=\left( s^{\mu },k_{\mu }\right) $ where the $s$'s
are the first $n+1$ $\hat{y}$'s and the $k$'s are the last $n+1$ $\hat{y}$'s
we have the following formula for the connection $D$ acting on them\footnote{%
Note that the indices go from $1$ to $2n+2$ and are different from the $2n$
operators defined above by $\left( \tilde{s}^{1},\ldots ,\tilde{s}^{n},%
\tilde{k}_{1},\ldots ,\tilde{k}_{n}\right) $. The difference between them is
the same as the difference between the embedding coordinates $\left(
x^{1},\ldots ,x^{n+1},p_{1},\ldots ,p_{n+1}\right) $ and $\left( \tilde{x}%
^{1},\ldots ,\tilde{x}^{n},\tilde{p}_{1},\ldots ,\tilde{p}_{n}\right) $.}
which is just plugging $\left( \text{%
%TCIMACRO{\TeXButton{\ref{D}}{\ref{D}}}%
%BeginExpansion
\ref{D}%
%EndExpansion
}\right) $ and $\left( \text{%
%TCIMACRO{\TeXButton{\ref{D2}}{\ref{D2}}}%
%BeginExpansion
\ref{D2}%
%EndExpansion
}\right) $ into the equation $\left( \text{%
%TCIMACRO{\TeXButton{\ref{Dy}}{\ref{Dy}}}%
%BeginExpansion
\ref{Dy}%
%EndExpansion
}\right) $ in 
%TCIMACRO{%
%\TeXButton{\hyperlink{step 2}{\textbf{step 2}}}{\hyperlink{step 2}{\textbf{step 2}}}}%
%BeginExpansion
\hyperlink{step 2}{\textbf{step 2}}%
%EndExpansion
:%
\begin{equation}
Ds^{\mu }=-\Gamma _{~\sigma \nu }^{\mu }dx^{\nu }s^{\sigma }  \label{Dyhat}
\end{equation}%
\begin{equation}
Dk_{\mu }:=-\frac{4}{3}R_{~(\mu \sigma )\beta }^{\psi }dx^{\beta }s^{\sigma
}p_{\psi }+\Gamma _{~\mu \sigma }^{\nu }dx^{\sigma }k_{\nu }  \notag
\end{equation}%
\begin{equation}
D^{2}s^{\mu }=dx^{\psi }dx^{\sigma }R_{~\nu \psi \sigma }^{\mu }s^{\nu }
\label{D2yhat}
\end{equation}%
\begin{equation}
D^{2}k_{\mu }=\frac{4}{3}dx^{\sigma }\left( C_{~\mu \beta \nu \sigma }^{\psi
}p_{\psi }dx^{\nu }+R_{~(\mu \beta )\sigma }^{\nu }\alpha _{\nu }\right)
s^{\beta }-R_{~\mu \sigma \beta }^{\nu }dx^{\sigma }dx^{\beta }k_{\nu } 
\notag
\end{equation}%
where again $C_{~abes}^{c}:=\nabla _{s}R_{~(ab)e}^{c}$.

\noindent \underline{Introducing terminology:}

In this paper when we say $f$ is a function/form we define it to be a
complex Taylor series in its variables\footnote{%
The set of all of these type of functions is sometimes called the enveloping
algebra of its arguments.}. Explicitly:%
\begin{equation}
f\left( u,\ldots ,v\right) =\sum_{l,j}f_{j_{1}\cdots j_{l}}u^{j_{1}}\cdots
v^{j_{l}}\text{ \ \ (}j\text{'s are powers not indices)}  \notag
\end{equation}%
where $u$ and $v$ are arbitrary.

So if $f$ is a function/form of some subset or all of the quantities $%
x,p,dx,dp,\omega ,\hbar $ and $i$ it then commutes with the $\hat{y}$'s and
will be called a complex-valued function/form. On the contrary an
matrix-valued function/form is a complex Taylor series in $\hat{y}$ and
possibly some subset or all of the quantities $x,p,dx,dp,\omega ,\hbar $ and 
$i$.

So if $f\left( x,p,dx,dp,\omega ,\hbar ,i\right) $ is a complex-valued
function/form it then commutes with the $\hat{y}$'s. More explicitly with
the matrix indices written (which are exceptions to our index conventions):%
\begin{equation*}
\left( \hat{y}^{A}\hat{y}^{B}\right) _{jk}=\sum_{l}\hat{y}_{jl}^{A}\hat{y}%
_{lk}^{B}
\end{equation*}%
\begin{equation*}
\left( \left[ \hat{y}^{A},f\right] \right) _{jk}:=\hat{y}_{jk}^{A}f-f\hat{y}%
_{jk}^{A}=0
\end{equation*}%
On the contrary a matrix-valued function/form does not. From now on we will
not write the matrix indices explicitly.

\noindent \underline{End Goal:}

The idea for Fedosov's introduction of the $\hat{y}$'s is to associate to
each $f\left( x,p\right) \in C^{\infty }\left( T^{\ast }M\right) $ a \textit{%
unique} observable $\hat{f}\left( x,p,\hat{y}\right) $: 
\begin{equation}
\hat{f}\left( x,p,\hat{y}\right) =\sum_{j,l}f_{j,l,A_{1}\cdots A_{l}}\hbar
^{j}\hat{y}^{A_{1}}\cdots \hat{y}^{A_{l}}  \notag
\end{equation}%
\textbf{Important Note:} Most of the rest of the sections will be dedicated
to finding an $\hat{f}$ (i.e., the coefficients $f_{j,l,A_{1}\cdots A_{l}}$)
for each $f\left( x,p\right) \in C^{\infty }\left( T^{\ast }M\right) $.

\subsubsection{The Constant Curvature Case Explicitly}

Specifically for $T^{\ast }M_{C_{p,q}}$ we have the induced symplectic form $%
\omega $ of $T^{\ast }%
%TCIMACRO{\U{211d} }%
%BeginExpansion
\mathbb{R}
%EndExpansion
^{n+1}$ onto $T^{\ast }M_{C_{p,q}}$ being: 
\begin{equation*}
\omega =\alpha _{\mu }dx^{\mu }=\left( \delta _{\mu }^{\nu }-Cx_{\mu }x^{\nu
}\right) \alpha _{\mu }dx^{\nu }
\end{equation*}%
From the definition of $\hat{y}$ the commutation relations in $\left( \text{%
%TCIMACRO{\TeXButton{\ref{yy}}{\ref{yy}}}%
%BeginExpansion
\ref{yy}%
%EndExpansion
}\right) $ and from the formula $\left( \text{%
%TCIMACRO{\TeXButton{\ref{g}}{\ref{g}}}%
%BeginExpansion
\ref{g}%
%EndExpansion
}\right) $:%
\begin{equation*}
\left[ s^{\mu },s^{\nu }\right] =0=\left[ k_{\mu },k_{\nu }\right] \ ,\ %
\left[ s^{\mu },k_{\nu }\right] =i\hbar \left( \delta _{\nu }^{\mu }-Cx^{\mu
}x_{\nu }\right)
\end{equation*}%
Since $dx^{\mu }$ and $\alpha _{\mu }$ are perpendicular to $x$ the matrix
counterparts $s^{\mu }$ and $k_{\mu }$ are also:%
\begin{equation}
\eta _{\mu \nu }x^{\mu }s^{\nu }=x^{\mu }k_{\mu }=0  \label{cond_sk}
\end{equation}%
Since $\eta _{\mu \nu }x^{\mu }s^{\nu }=x^{\mu }k_{\mu }=0$ we have $n$
independent operators which is required since (one for each direction on $%
M_{C_{p,q}}$).

The action of the connection and curvature acting on $s^{\mu }$ \& $k_{\mu }$
on a general phase-space and not just $T^{\ast }M_{C_{p,q}}$ is written down
directly using the formulas in $\left( \text{%
%TCIMACRO{\TeXButton{\ref{g}}{\ref{g}}}%
%BeginExpansion
\ref{g}%
%EndExpansion
}\right) $ into the formula $\left( \text{%
%TCIMACRO{\TeXButton{\ref{Dyhat}}{\ref{Dyhat}}}%
%BeginExpansion
\ref{Dyhat}%
%EndExpansion
}\right) $.

\subsection{Constructing the Global Derivation}

In 
%TCIMACRO{%
%\TeXButton{\hyperlink{step 3}{\textbf{step 3}}}{\hyperlink{step 3}{\textbf{step 3}}} }%
%BeginExpansion
\hyperlink{step 3}{\textbf{step 3}}
%EndExpansion
in the algorithm, must determine a global derivation as a matrix commutator $%
\hat{D}=\left[ \hat{Q},\cdot \right] $ which is central to constructing the
coefficients $f_{A_{1}\cdots A_{l}}$ in equation $\left( \text{%
%TCIMACRO{\TeXButton{\ref{cond_fhat}}{\ref{cond_fhat}}}%
%BeginExpansion
\ref{cond_fhat}%
%EndExpansion
}\right) $ for each $f\left( x,p\right) \in C^{\infty }\left( T^{\ast
}M\right) $.

Define the derivation $\hat{D}$ by the graded commutator:%
\begin{equation*}
\hat{D}=\left[ \hat{Q},\cdot \right] /i\hbar =\left[ \hat{Q}_{A}\Theta
^{A},\cdot \right] /i\hbar
\end{equation*}%
\begin{equation*}
\hat{Q}_{A}=\sum_{l}Q_{AA_{1}\cdots A_{l}}\hat{y}^{A_{1}}\cdots \hat{y}%
^{A_{l}}
\end{equation*}%
where $\Theta ^{B}=\left( dx^{\sigma },\alpha _{\sigma }\right) $ (see again
the definition for $\alpha $ in $\left( \text{%
%TCIMACRO{\TeXButton{\ref{alpha}}{\ref{alpha}}}%
%BeginExpansion
\ref{alpha}%
%EndExpansion
}\right) $) and $Q_{AA_{1}\cdots A_{l}}$\ are complex-valued functions of $x$
and $p$ that need to be determined. We reiterate that complex-valued
functions are\ not matrices hence they commute with the $\hat{y}$'s.

In 
%TCIMACRO{%
%\TeXButton{\hyperlink{step 3}{\textbf{step 3}}}{\hyperlink{step 3}{\textbf{step 3}}} }%
%BeginExpansion
\hyperlink{step 3}{\textbf{step 3}}
%EndExpansion
we have the mysterious condition $\left( \text{%
%TCIMACRO{\TeXButton{\ref{cond_Dhat}}{\ref{cond_Dhat}}}%
%BeginExpansion
\ref{cond_Dhat}%
%EndExpansion
}\right) $ that partially determines the functions $Q_{AA_{1}\cdots A_{l}}$:

We rewrite the condition $\left( \text{%
%TCIMACRO{\TeXButton{\ref{cond_Dhat}}{\ref{cond_Dhat}}}%
%BeginExpansion
\ref{cond_Dhat}%
%EndExpansion
}\right) $ as:%
\begin{equation}
\left( D-\hat{D}\right) ^{2}\hat{y}^{A}=\left[ \Omega -D\hat{Q}+\hat{Q}%
^{2}/i\hbar ,\hat{y}^{A}\right] /i\hbar =0  \label{DDhat2}
\end{equation}%
where $\Omega $ is the phase-space curvature as a commutator (see $\left[ 
\text{Fed}\right] $ for the details):%
\begin{equation}
\frac{1}{i\hbar }\left[ \Omega ,\hat{y}^{A}\right] :=D^{2}\hat{y}^{A}=R_{B}^{%
\text{ \ }A}\hat{y}^{B}  \label{Omega}
\end{equation}%
with solution $\Omega :=-\frac{1}{2}\omega _{AC}R_{CEB}^{\text{ \ ~~~~}%
A}\Theta ^{C}\wedge \Theta ^{E}\hat{y}^{B}\hat{y}^{C}$ where $R_{CEB}^{\text{
\ ~~~~}A}$ is the phase-space curvature.

From now on we let\footnote{%
This is the same as the condition of Fedosov $\Omega -Dr+\hat{d}r+r^{2}=0$.
See 
%TCIMACRO{\TeXButton{Fedosov B. 1996}{\hyperlink{ref1}{Fedosov B. 1996}}}%
%BeginExpansion
\hyperlink{ref1}{Fedosov B. 1996}%
%EndExpansion
, and 
%TCIMACRO{%
%\TeXButton{Gadella M. et al 2005}{\hyperlink{ref1}{Gadella M. \textit{et al} 2005}}}%
%BeginExpansion
\hyperlink{ref1}{Gadella M. \textit{et al} 2005}%
%EndExpansion
.}: 
\begin{equation}
\Omega -D\hat{Q}+\hat{Q}^{2}/i\hbar =0  \label{Qhat}
\end{equation}%
and keep it in the back of our minds that we could add something that
commutes with all $\hat{y}$'s to $\Omega -D\hat{Q}+\hat{Q}^{2}/i\hbar $.

To emphasize the importance of this equation the reader should note that the
whole Fedosov $\ast $ hinges on this $\hat{Q}$ existing. We know a solution
exists perturbatively in general (the recursive solution for it is in 
%TCIMACRO{\TeXButton{Fedosov B. 1996}{\hyperlink{ref1}{Fedosov B. 1996}} }%
%BeginExpansion
\hyperlink{ref1}{Fedosov B. 1996}
%EndExpansion
on p. 144), however, convergence issues of the general series still remain.
We have found that solving for $\hat{Q}$ to be the hardest point of the
computation of the Fedosov $\ast $ because of the need for the right ansatz
to the nonlinear equation $\left( \text{%
%TCIMACRO{\TeXButton{\ref{Qhat}}{\ref{Qhat}}}%
%BeginExpansion
\ref{Qhat}%
%EndExpansion
}\right) $.

Fedosov at this point would implement an algorithm to construct $\hat{Q}$
perturbatively, however, rather than do this we will make an ansatz for $%
\hat{Q}$ using some ingenuity. This will give us an exact solution for $\hat{%
Q}$.

$\Omega $ is:%
\begin{eqnarray}
\Omega &=&-R_{~\mu \sigma \beta }^{\nu }dx^{\sigma }dx^{\beta }k_{\nu
}s^{\mu }+\frac{2}{3}D\left( R_{~(\mu \beta )\sigma }^{\nu }p_{\nu }s^{\beta
}s^{\mu }dx^{\sigma }\right)  \label{Omega_general} \\
&=&-R_{~\mu \sigma \beta }^{\nu }dx^{\sigma }dx^{\beta }k_{\nu }s^{\mu }+%
\frac{2}{3}dx^{\sigma }\left( C_{~\mu \beta \nu \sigma }^{\psi }p_{\psi
}dx^{\nu }+R_{~(\mu \beta )\sigma }^{\nu }\alpha _{\nu }\right) s^{\beta
}s^{\mu }  \notag
\end{eqnarray}

where $C_{~abes}^{c}:=\nabla _{s}R_{~(ab)e}^{c}$.

We verify that it gives the curvature as commutators:%
\begin{equation*}
\frac{1}{i\hbar }\left[ \Omega ,s^{\mu }\right] =D^{2}s^{\mu }=R_{~\nu \psi
\varepsilon }^{\mu }dx^{\psi }dx^{\varepsilon }s^{\nu }
\end{equation*}%
\begin{equation*}
\frac{1}{i\hbar }\left[ \Omega ,k_{\mu }\right] =D^{2}k_{\mu }=\frac{4}{3}%
dx^{\sigma }\left( C_{~\mu \beta \nu \sigma }^{\psi }p_{\psi }dx^{\nu
}+R_{~(\mu \beta )\sigma }^{\nu }\alpha _{\nu }\right) s^{\beta }-R_{~\mu
\sigma \beta }^{\nu }dx^{\sigma }dx^{\beta }k_{\nu }
\end{equation*}%
Our ansatz for a solution to the equation $\left( \text{%
%TCIMACRO{\TeXButton{\ref{Qhat}}{\ref{Qhat}}}%
%BeginExpansion
\ref{Qhat}%
%EndExpansion
}\right) $ is:%
\begin{eqnarray}
\hat{Q} &=&\left( s^{\mu }\alpha _{\mu }-z_{\mu }dx^{\mu }\right) +j^{\mu
}\alpha _{\mu }+z_{\nu }f_{~\mu }^{\nu }dx^{\mu }  \label{Qhat ansatz} \\
&&+p_{\nu }\left( \left( D+f_{~\rho }^{\sigma }dx^{\rho }\hat{\partial}%
_{\sigma }-dx^{\sigma }\hat{\partial}_{\sigma }\right) j^{\nu }+\Gamma
_{~\rho \sigma }^{\nu }dx^{\sigma }j^{\rho }-\frac{2}{3}R_{~(\mu \beta
)\sigma }^{\nu }s^{\beta }s^{\mu }dx^{\sigma }\right)  \notag
\end{eqnarray}%
where $\hat{\partial}_{\mu }:=\partial /\partial s^{\mu }$ and along with
condition on $f_{~\mu }^{\nu }$:%
\begin{equation}
\left( \left( D+f_{~\rho }^{\mu }dx^{\rho }\hat{\partial}_{\mu }-dx^{\mu }%
\hat{\partial}_{\mu }\right) f_{~\sigma }^{\nu }+\Gamma _{~\rho \mu }^{\nu
}dx^{\mu }f_{~\sigma }^{\rho }-\Gamma _{~\sigma \mu }^{\nu }dx^{\mu
}+R_{~\mu \beta \sigma }^{\nu }s^{\mu }dx^{\beta }\right) dx^{\sigma }=0
\label{cond Qhat}
\end{equation}%
To see that the term:%
\begin{equation*}
p_{\nu }\left( \left( D+f_{~\rho }^{\sigma }dx^{\rho }\hat{\partial}_{\sigma
}-\hat{\partial}_{c}\right) j^{\nu }+\Gamma _{~\rho \sigma }^{\nu
}dx^{\sigma }j^{\rho }\right)
\end{equation*}%
in $\left( \text{%
%TCIMACRO{\TeXButton{\ref{Qhat ansatz}}{\ref{Qhat ansatz}}}%
%BeginExpansion
\ref{Qhat ansatz}%
%EndExpansion
}\right) $ is coordinate independent if $j^{a}$ and $f_{~c}^{e}$ are we
express it in terms of abstract indices:%
\begin{equation*}
p_{b}\left( \nabla _{c}j^{b}+f_{~c}^{e}\hat{\partial}_{e}-\hat{\partial}%
_{c}\right) j^{b}
\end{equation*}%
where we used the fact that $Dj^{b}=\Theta ^{C}D_{C}j^{b}=\nabla _{c}j^{b}$
because $j^{b}$ is a function of $x$ and $s$ only. So we can see that if $%
j^{a}$ and $f_{~c}^{e}$ are independent of the choice of configuration space
coordinates then so is $\hat{Q}$.

By putting $\left( \text{%
%TCIMACRO{\TeXButton{\ref{Qhat ansatz}}{\ref{Qhat ansatz}}}%
%BeginExpansion
\ref{Qhat ansatz}%
%EndExpansion
}\right) $ and $\left( \text{%
%TCIMACRO{\TeXButton{\ref{cond Qhat}}{\ref{cond Qhat}}}%
%BeginExpansion
\ref{cond Qhat}%
%EndExpansion
}\right) $ into\ the equation $\left( \text{%
%TCIMACRO{\TeXButton{\ref{Qhat}}{\ref{Qhat}}}%
%BeginExpansion
\ref{Qhat}%
%EndExpansion
}\right) $ and performing a straightforward calculation we can easily verify
that they solve the equation in $\left( \text{%
%TCIMACRO{\TeXButton{\ref{Qhat}}{\ref{Qhat}}}%
%BeginExpansion
\ref{Qhat}%
%EndExpansion
}\right) $. Moreover, the equation $\left( \text{%
%TCIMACRO{\TeXButton{\ref{cond Qhat}}{\ref{cond Qhat}}}%
%BeginExpansion
\ref{cond Qhat}%
%EndExpansion
}\right) $ is locally integrable for $f_{~\mu }^{\nu }$ by the
Cauchy-Kovalevskaya theorem (see 
%TCIMACRO{%
%\TeXButton{\hyperlink{Appendix B}{appendix B}}{\hyperlink{Appendix B}{appendix B}}}%
%BeginExpansion
\hyperlink{Appendix B}{appendix B}%
%EndExpansion
). This fact allows us to come up with an iterative solution in the spirit
of the series of Fedosov star-product.

We have therefore proved the following theorem:

\bigskip

\noindent \textbf{Thm.} Given any cotangent bundle $T^{\ast }M$, \ the
solution to the equation in $\left( \text{%
%TCIMACRO{\TeXButton{\ref{Qhat}}{\ref{Qhat}}}%
%BeginExpansion
\ref{Qhat}%
%EndExpansion
}\right) $ is $\left( \text{%
%TCIMACRO{\TeXButton{\ref{Qhat ansatz}}{\ref{Qhat ansatz}}}%
%BeginExpansion
\ref{Qhat ansatz}%
%EndExpansion
}\right) $ along with the condition in $\left( \text{%
%TCIMACRO{\TeXButton{\ref{cond Qhat}}{\ref{cond Qhat}}}%
%BeginExpansion
\ref{cond Qhat}%
%EndExpansion
}\right) $ where the equation $\left( \text{%
%TCIMACRO{\TeXButton{\ref{cond Qhat}}{\ref{cond Qhat}}}%
%BeginExpansion
\ref{cond Qhat}%
%EndExpansion
}\right) $ is locally integrable for $f_{~\mu }^{\nu }$ by the
Cauchy-Kovalevskaya theorem.

\subsubsection{The Constant Curvature Case Explicitly}

The solution that was found for our example of $T^{\ast }M_{C_{p,q}}$ using
the above ansatz $\left( \text{%
%TCIMACRO{\TeXButton{\ref{Qhat ansatz}}{\ref{Qhat ansatz}}}%
%BeginExpansion
\ref{Qhat ansatz}%
%EndExpansion
}\right) $ and condition $\left( \text{%
%TCIMACRO{\TeXButton{\ref{cond Qhat}}{\ref{cond Qhat}}}%
%BeginExpansion
\ref{cond Qhat}%
%EndExpansion
}\right) $:%
\begin{equation}
\hat{Q}=\left( s^{\mu }\alpha _{\mu }-z_{\mu }dx^{\mu }\right) -C\left(
z_{\nu }s^{\nu }\right) \left( s_{\mu }dx^{\mu }\right) +\frac{C}{3}\left(
\left( p_{\nu }s^{\nu }\right) \left( s_{\mu }dx^{\mu }\right) -\left(
p_{\nu }dx^{\nu }\right) u\right)  \label{Qhat soln}
\end{equation}%
where $z_{\mu }:=k_{\mu }+p_{\mu }$, $u=\eta _{\mu \nu }s^{\mu }s^{\nu }$
and $p_{\mu }x^{\mu }=\eta _{\mu \nu }s^{\mu }x^{\nu }=k_{\mu }x^{\mu
}=\alpha _{\mu }x^{\mu }=\eta _{\mu \nu }dx^{\mu }x^{\nu }=0$.

\subsection{The Basis For the Algebra of Observables}

Now we have all the tools in place to associate an observable $\hat{f}$ to
every $f\in C^{\infty }\left( T^{\ast }M\right) $. At 
%TCIMACRO{%
%\TeXButton{\hyperlink{step 4}{\textbf{step 4}}}{\hyperlink{step 4}{\textbf{step 4}}} }%
%BeginExpansion
\hyperlink{step 4}{\textbf{step 4}}
%EndExpansion
in our algorithm we require that every observable $\hat{f}\left( x,p,\hat{y}%
\right) $ must satisfy the equation $\left( \text{%
%TCIMACRO{\TeXButton{\ref{cond_fhat}}{\ref{cond_fhat}}}%
%BeginExpansion
\ref{cond_fhat}%
%EndExpansion
}\right) $.

The condition $f_{j,l,(A_{1}\cdots A_{l})}=f_{j,l,A_{1}\cdots A_{l}}$ is the
condition for Weyl or symmetric quantization. You can choose another
ordering, but this is sufficient. Moreover, this is the choice that Fedosov
makes for ordering. The condition in $\left( \text{%
%TCIMACRO{\TeXButton{\ref{cond_fhat}}{\ref{cond_fhat}}}%
%BeginExpansion
\ref{cond_fhat}%
%EndExpansion
}\right) $ is used to solve for a unique $\hat{f}$ to every $f\in C^{\infty
}\left( T^{\ast }M\right) $ up to some "reasonable" ambiguity.

Here we (again) diverge from the Fedosov algorithm. Instead of constructing
the coefficients so that $f_{j,l,A_{1}\cdots A_{l}}$ is symmetric in $\hat{y}
$'s we instead require that each term in:%
\begin{equation}
\hat{f}\left( \hat{x},\hat{p}\right) =\sum_{jlm}\tilde{f}_{j,l,m,\mu
_{1}\cdots \mu _{l}}^{\nu _{1}\cdots \nu _{m}}\hbar ^{j}\hat{x}^{\mu
_{1}}\cdots \hat{x}^{\mu _{l}}\hat{p}_{\nu _{1}}\cdots \hat{p}_{\nu _{m}}
\label{soln fhat}
\end{equation}%
where $\tilde{f}_{j,l,m,\mu _{1}\cdots \mu _{l}}^{\nu _{1}\cdots \nu _{m}}$
is a complex-valued function of $x$ and $p$ and is symmetric in all $\hat{x}$
and $\hat{p}$, i.e.:%
\begin{equation*}
\hat{f}\left( \hat{x},\hat{p}\right) =\sum_{lm}\tilde{f}_{j,l,m,\mu
_{1}\cdots \mu _{l}}^{\nu _{1}\cdots \nu _{m}}\hbar ^{j}SYM\left( \hat{x}%
^{\mu _{1}}\cdots \hat{x}^{\mu _{l}}\hat{p}_{\nu _{1}}\cdots \hat{p}_{\nu
_{m}}\right) 
\end{equation*}%
where:%
\begin{eqnarray*}
SYM\left( \hat{x}^{\mu _{1}}\cdots \hat{x}^{\mu _{l}}\hat{p}_{\nu
_{1}}\cdots \hat{p}_{\nu _{m}}\right)  &=&\hat{x}^{\mu _{1}}\cdots \hat{x}%
^{\mu _{l}}\hat{p}_{\nu _{1}}\cdots \hat{p}_{\nu _{m}}+\left( \text{all
perms. of }\hat{x}\text{'s and }\hat{p}\text{'s}\right)  \\
&=&\hat{x}^{\mu _{1}}\cdots \hat{x}^{\mu _{l}}\hat{p}_{\nu _{1}}\cdots \hat{p%
}_{\nu _{m}}+\hat{x}^{\mu _{1}}\cdots \hat{x}^{\mu _{l-1}}\hat{p}_{\nu _{1}}%
\hat{x}^{\mu _{l}}\hat{p}_{\nu _{2}}\cdots \hat{p}_{\nu _{m}}+\cdots 
\end{eqnarray*}%
The definition of $\hat{f}$ in $\left( \text{%
%TCIMACRO{\TeXButton{\ref{soln fhat}}{\ref{soln fhat}}}%
%BeginExpansion
\ref{soln fhat}%
%EndExpansion
}\right) $ corresponds to the phase-space function:%
\begin{equation*}
\sigma \left( \hat{f}\right) =f\left( x,p\right) =\sum_{jlm}\tilde{f}%
_{j,l,m,\mu _{1}\cdots \mu _{l}}^{\nu _{1}\cdots \nu _{m}}\hbar ^{j}x^{\mu
_{1}}\cdots x^{\mu _{l}}p_{\nu _{1}}\cdots p_{\nu _{m}}
\end{equation*}%
The nice property of the above form of $\left( \text{%
%TCIMACRO{\TeXButton{\ref{soln fhat}}{\ref{soln fhat}}}%
%BeginExpansion
\ref{soln fhat}%
%EndExpansion
}\right) $ is that the coefficients $\tilde{f}_{j,l,m,\mu _{1}\cdots \mu
_{l}}^{\nu _{1}\cdots \nu _{m}}$ are constant. This is easily see by acting $%
\left( D-\hat{D}\right) $\ on the equation. Also, the formula is nice
because now we can find any basis $\left( \hat{x},\hat{p}\right) $ and these
will give unique $\hat{f}$ for all phase-space function $f$. All we need to
do now is find any basis $\left( \hat{x},\hat{p}\right) $ which is our next
task.\pagebreak 

\noindent \underline{Finding a Basis:}

We define a basis $\left( \hat{x},\hat{p}\right) $ as any operator of the
form:%
\begin{equation}
\hat{x}^{\mu }=\sum_{l}b_{j,l,A_{1}\cdots A_{l}}^{\mu }\hbar ^{j}\hat{y}%
^{A_{1}}\cdots \hat{y}^{A_{l}}  \label{xhat}
\end{equation}%
\begin{equation}
\hat{p}_{\mu }=\sum_{l}c_{j,l,\mu ,A_{1}\cdots A_{l}}\hbar ^{j}\hat{y}%
^{A_{1}}\cdots \hat{y}^{A_{l}}  \label{phat}
\end{equation}%
where $b_{j,l,A_{1}\cdots A_{l}}^{\mu }$\ and $c_{j,l,\mu ,A_{1}\cdots
A_{l}} $ are complex-valued functions of $x$ and $p$ (which are the
coefficients $f_{j,l,A_{1}\cdots A_{l}}$ in equation $\left( \text{%
%TCIMACRO{\TeXButton{\ref{cond_fhat}}{\ref{cond_fhat}}}%
%BeginExpansion
\ref{cond_fhat}%
%EndExpansion
}\right) $ where $f=x$ or $f=p\,$\ respectively) and will be partially
determined by the equations:%
\begin{equation}
\left( D-\hat{D}\right) \hat{x}^{\mu }=0~~~,~~\ \sigma \left( \hat{x}^{\mu
}\right) =b_{0,0}^{\mu }=x^{\mu }  \label{cond xhat}
\end{equation}%
\begin{equation}
\left( D-\hat{D}\right) \hat{p}_{\mu }=0~~~,~~~\sigma \left( \hat{p}_{\mu
}\right) =c_{0,0,\mu }=p_{\mu }  \label{cond phat}
\end{equation}%
Remember that our observables are defined in $\left( \text{%
%TCIMACRO{\TeXButton{\ref{fhat}}{\ref{fhat}}}%
%BeginExpansion
\ref{fhat}%
%EndExpansion
}\right) $. To express them in the form of $\left( \text{%
%TCIMACRO{\TeXButton{\ref{soln fhat}}{\ref{soln fhat}}}%
%BeginExpansion
\ref{soln fhat}%
%EndExpansion
}\right) $ we need to invert the relations $\left( \text{%
%TCIMACRO{\TeXButton{\ref{cond xhat}}{\ref{cond xhat}}}%
%BeginExpansion
\ref{cond xhat}%
%EndExpansion
}\right) $ and $\left( \text{%
%TCIMACRO{\TeXButton{\ref{cond phat}}{\ref{cond phat}}}%
%BeginExpansion
\ref{cond phat}%
%EndExpansion
}\right) $ so that we express $\hat{y}$ in terms of $x$, $p$, $\hat{x}$, and 
$\hat{p}$ as a the matrix-valued function $\hat{y}^{A}=\hat{y}^{A}\left( x,p,%
\hat{x},\hat{p}\right) $. By substituting $\hat{y}^{A}=\hat{y}^{A}\left( x,p,%
\hat{x},\hat{p}\right) $ into $\left( \text{%
%TCIMACRO{\TeXButton{\ref{fhat}}{\ref{fhat}}}%
%BeginExpansion
\ref{fhat}%
%EndExpansion
}\right) $ it will be observed that all observables can be expressed in the
form of $\left( \text{%
%TCIMACRO{\TeXButton{\ref{soln fhat}}{\ref{soln fhat}}}%
%BeginExpansion
\ref{soln fhat}%
%EndExpansion
}\right) $. Of course, the caveat is that we have assumed the convergence of
all of these series which will not be true in general.

To construct a basis $\left( \hat{x},\hat{p}\right) $ for the algebra
Fedosov at this point would implement an algorithm yielding perturbative
solutions (see 
%TCIMACRO{\TeXButton{Fedosov B. 1996}{\hyperlink{ref1}{Fedosov B. 1996}} }%
%BeginExpansion
\hyperlink{ref1}{Fedosov B. 1996}
%EndExpansion
p 146). We instead try to find exact solutions to them.\footnote{%
We, again, ran the Fedosov algorithm a few times to help us see what for the
ansatz should take.}

\subsubsection{The Constant Curvature Case Explicitly}

Specifically for the case of $T^{\ast }M_{C_{p,q}}$ we make the ansatz for
both $\hat{x}$ and $\hat{p}$:%
\begin{equation*}
\hat{x}^{\mu }=f\left( u\right) x^{\mu }+h\left( u\right) s^{\mu }
\end{equation*}%
\begin{equation*}
\hat{p}_{\mu }=z_{\nu }s^{\nu }x_{\mu }g\left( u\right) +z_{\mu }j\left(
u\right)
\end{equation*}%
where $u:=\eta _{\mu \nu }s^{\mu }s^{\nu }$ and $z_{\mu }:=k_{\mu }+p_{\mu }$%
.

We require that both $\hat{x}$ and $\hat{p}$ satisfy the two conditions $%
\left( \text{%
%TCIMACRO{\TeXButton{\ref{cond xhat}}{\ref{cond xhat}}}%
%BeginExpansion
\ref{cond xhat}%
%EndExpansion
}\right) $ and $\left( \text{%
%TCIMACRO{\TeXButton{\ref{cond phat}}{\ref{cond phat}}}%
%BeginExpansion
\ref{cond phat}%
%EndExpansion
}\right) $ and by solving the subsequent differential equations we obtain
the solutions:%
\begin{equation}
\hat{x}^{\mu }=\left( x^{\mu }+s^{\mu }\right) \frac{1}{\sqrt{Cu+1}}
\label{xhat_soln2}
\end{equation}%
\begin{equation}
\hat{p}_{\mu }=\left( -Cz_{\nu }s^{\nu }x_{\mu }+z_{\mu }\right) \sqrt{Cu+1}%
-iC\hbar n\hat{x}_{\mu }  \label{phat_soln2}
\end{equation}%
where $u=s_{\mu }s^{\mu }$, $z_{\mu }:=k_{\mu }+p_{\mu }$, and with the
computed conditions:%
\begin{equation*}
\sigma \left( \hat{x}^{\mu }\right) =b_{0,0}^{\mu }=x^{\mu }~~,~~~\sigma
\left( \hat{p}_{\mu }\right) =c_{0,0,\mu }=p_{\mu }
\end{equation*}%
\begin{equation}
\hat{x}\cdot \hat{x}=1/C~~~,~\ \ ~\hat{x}\cdot \hat{p}=\hat{p}\cdot \hat{x}%
-ni\hbar =0  \label{xhatphat conds2}
\end{equation}%
We now use these results to write the solution for $\hat{x}$ and $\widehat{%
\tilde{p}}$ for the embedding:

\begin{equation*}
x^{\mu }x_{\mu }=1/C~~\ ,~~\ x^{\mu }\tilde{p}_{\mu }=A
\end{equation*}%
Since this is a canonical transformation:%
\begin{equation*}
\tilde{p}_{\mu }=p_{\mu }+CAx_{\mu }~~~,~~~\tilde{x}^{\mu }=x^{\mu }
\end{equation*}%
and preserves all constraints except $x^{\mu }p_{\mu }=A$ we can write the
solution as:%
\begin{equation*}
\widehat{\tilde{p}}_{\mu }=\hat{p}_{\mu }+CA\hat{x}_{\mu }~~~,~~~\widehat{%
\tilde{x}}^{\mu }=\hat{x}^{\mu }
\end{equation*}%
(see 
%TCIMACRO{%
%\TeXButton{\hyperlink{Appendix C}{appendix C}}{\hyperlink{Appendix C}{appendix C}} }%
%BeginExpansion
\hyperlink{Appendix C}{appendix C}
%EndExpansion
for proof) so:%
\begin{equation*}
\widehat{\tilde{x}}^{\mu }=\hat{x}^{\mu }=\left( x^{\mu }+s^{\mu }\right) 
\frac{1}{\sqrt{Cu+1}}
\end{equation*}%
\begin{equation*}
\widehat{\tilde{p}}_{\mu }=\left( -Cz_{\nu }s^{\nu }x_{\mu }+z_{\mu }\right) 
\sqrt{Cu+1}+C\left( A-i\hbar n\right) \hat{x}_{\mu }
\end{equation*}%
\textbf{Note:} From now on we will use the embedding (and by dropping the
tilde):%
\begin{equation*}
x^{\mu }x_{\mu }=1/C~~\ ,~~\ x^{\mu }p_{\mu }=A
\end{equation*}%
and the solutions:%
\begin{equation}
\hat{x}^{\mu }=\left( x^{\mu }+s^{\mu }\right) \frac{1}{\sqrt{Cu+1}}
\label{xhat_soln2A}
\end{equation}%
\begin{equation}
\hat{p}_{\mu }=\left( -Cz_{\nu }s^{\nu }x_{\mu }+z_{\mu }\right) \sqrt{Cu+1}%
+C\left( A-i\hbar n\right) \hat{x}_{\mu }  \label{phat_soln2A}
\end{equation}%
with computed conditions:%
\begin{equation}
\hat{x}\cdot \hat{x}=1/C~~~,~\ \ ~\hat{x}\cdot \hat{p}=\hat{p}\cdot \hat{x}%
-ni\hbar =A  \label{xhatphat conds2A}
\end{equation}%
In group theoretic terminology the two conditions above represent the
Casimir invariants of the algebra of observables.

\subsection{The Commutators}

%TCIMACRO{\TeXButton{\label{s:The Commutators}}{\label{s:The Commutators}}}%
%BeginExpansion
\label{s:The Commutators}%
%EndExpansion
Once we have $\hat{x}^{\mu }$ and $\hat{p}_{\mu }$, i.e., the coefficients $%
b_{j,l,A_{1}\cdots A_{l}}^{\mu }$ and $c_{j,l,\mu ,A_{1}\cdots A_{l}}$\ we
work out the commutation relations $\left[ \hat{x}^{\mu },\hat{x}^{\nu }%
\right] ,\left[ \hat{x}^{\mu },\hat{p}_{\nu }\right] $ and $\left[ \hat{p}%
_{\mu },\hat{p}_{\nu }\right] $ using the solution for $\hat{x}$ and $\hat{p}
$ (for either case they are $\left( \text{%
%TCIMACRO{\TeXButton{\ref{xhat_soln2}}{\ref{xhat_soln2}}}%
%BeginExpansion
\ref{xhat_soln2}%
%EndExpansion
}\right) $ and $\left( \text{%
%TCIMACRO{\TeXButton{\ref{phat_soln2}}{\ref{phat_soln2}}}%
%BeginExpansion
\ref{phat_soln2}%
%EndExpansion
}\right) $) in a brute force calculation:%
\begin{equation*}
\hat{h}\left( \hat{x},\hat{p}\right) :=\left[ \hat{f}\left( \hat{x},\hat{p}%
\right) ,\hat{g}\left( \hat{x},\hat{p}\right) \right]
\end{equation*}%
\begin{equation}
\implies \left[ f_{\ast }\left( x,p\right) ,g_{\ast }\left( x,p\right) %
\right] _{\ast }=h_{\ast }\left( x,p\right) =i\hbar \left[ f,g\right]
_{P}+O\left( \hbar ^{2}\right)  \label{*-comm}
\end{equation}%
where $\hat{f}$, $\hat{g}$, $\hat{h}$ and $f_{\ast }$, $g_{\ast }$, $h_{\ast
}$ are functions defined by:%
\begin{equation*}
\hat{f}\left( \hat{x},\hat{p}\right) =\sum_{lmj}\tilde{f}_{j,l,m,\mu
_{1}\cdots \mu _{l}}^{\nu _{1}\cdots \nu _{m}}\hbar ^{j}\hat{x}^{\mu
_{1}}\cdots \hat{x}^{\mu _{l}}\hat{p}_{\nu _{1}}\cdots \hat{p}_{\nu _{m}}
\end{equation*}%
\begin{equation*}
f_{\ast }\left( x,p\right) =\sum_{lmj}\tilde{f}_{j,l,m,\mu _{1}\cdots \mu
_{l}}^{\nu _{1}\cdots \nu _{m}}\hbar ^{j}x^{\mu _{1}}\ast \cdots \ast x^{\mu
_{l}}\ast p_{\nu _{1}}\ast \cdots \ast p_{\nu _{m}}
\end{equation*}%
where $\tilde{f}_{j,l,m,\mu _{1}\cdots \mu _{l}}^{\nu _{1}\cdots \nu _{m}}$
are constants.

These two sets, one of all $f_{\ast }$'s $\left\{ f_{\ast }\right\} $ and
one of all $\hat{f}$'s $\left\{ \hat{f}\right\} $ defined above are
isomorphic with isomorphism $\sigma ^{-1}$.

\subsubsection{The Constant Curvature Case Explicitly}

In our case of $T^{\ast }M_{C_{p,q}}$ we compute:%
\begin{equation}
\left[ \hat{x}^{\mu },\hat{x}^{\nu }\right] =0  \label{commutators_general}
\end{equation}%
\begin{equation*}
\left[ \hat{x}^{\mu },\hat{p}_{\nu }\right] =i\hbar \left( \delta _{\nu
}^{\mu }-C\hat{x}^{\mu }\hat{x}_{\nu }\right)
\end{equation*}%
\begin{equation*}
\left[ \hat{p}_{\mu },\hat{p}_{\nu }\right] =2i\hbar C\hat{x}_{[\nu }\hat{p}%
_{\mu ]}
\end{equation*}%
along with the computed conditions:%
\begin{equation*}
\hat{x}^{\mu }\hat{x}_{\mu }=1/C~~~,~\ \ ~\hat{p}_{\mu }\hat{x}^{\mu
}+ni\hbar =\hat{x}^{\mu }\hat{p}_{\mu }=A
\end{equation*}%
We now define:%
\begin{equation*}
\hat{M}_{\mu \nu }=\hat{x}_{[\mu }\hat{p}_{\nu ]}=\hat{p}_{[\nu }\hat{x}%
_{\mu ]}=\left( -Cz_{\rho }s^{\rho }x_{[\nu }+z_{[\nu }\right) \left( x_{\mu
]}+s_{\mu ]}\right)
\end{equation*}%
The leading order term is found to be:%
\begin{equation*}
\sigma \left( \hat{M}_{\mu \nu }\right) =x_{[\mu }p_{\nu ]}=M_{\mu \nu }
\end{equation*}%
We recognize that $\hat{M}$ and $\hat{x}$ are the more "natural" variables
than $\hat{x}$ and $\hat{p}$ because $\hat{p}_{\mu }\hat{x}^{\mu }=-ni\hbar $
and $\hat{x}^{\mu }\hat{p}_{\mu }=A$ where $A$ is an arbitrary constant.
These are very "unnatural" since there is no reason why it shouldn't be $%
\hat{p}_{\mu }\hat{x}^{\mu }=A$ and $\hat{x}^{\mu }\hat{p}_{\mu }=ni\hbar $
or something else like this. $\hat{M}$ projects out the part of the momentum 
$\hat{p}$ that is parallel to $\hat{x}$ ($2\hat{x}^{\mu }\hat{M}_{\mu \nu }=%
\hat{p}_{\nu }/C-A\hat{x}_{\nu }$). We regard this part of $\hat{p}$ to be
irrelevant because it does not affect the form of the commutators in $\left( 
\text{%
%TCIMACRO{\TeXButton{\ref{commutators_general}}{\ref{commutators_general}}}%
%BeginExpansion
\ref{commutators_general}%
%EndExpansion
}\right) $ and it preserves the symplectic form.

We have the definitions:%
\begin{equation}
\hat{x}^{\mu }=\left( x^{\mu }+s^{\mu }\right) \frac{1}{\sqrt{Cu+1}}
\label{xhatMhat solns}
\end{equation}%
\begin{equation*}
\hat{M}_{\mu \nu }=\hat{x}_{[\mu }\hat{p}_{\nu ]}=\hat{p}_{[\nu }\hat{x}%
_{\mu ]}=-Cz_{\rho }s^{\rho }x_{[\nu }s_{\mu ]}+z_{[\nu }x_{\mu ]}+z_{[\nu
}s_{\mu ]}
\end{equation*}

\noindent and the computed commutation relations (which is again very
straightforward):%
\begin{equation}
\left[ \hat{x}^{\mu },\hat{x}^{\nu }\right] =0  \label{comm xhatMhat}
\end{equation}%
\begin{equation*}
\left[ \hat{x}_{\mu },\hat{M}_{\nu \rho }\right] =i\hbar \hat{x}_{[\nu }\eta
_{\rho ]\mu }
\end{equation*}%
\begin{equation*}
\left[ \hat{M}_{\mu \nu },\hat{M}_{\rho \sigma }\right] =i\hbar \left( \hat{M%
}_{\sigma \lbrack \mu }\eta _{\nu ]\rho }-\hat{M}_{\rho \lbrack \mu }\eta
_{\nu ]\sigma }\right)
\end{equation*}%
subject to the conditions:%
\begin{equation}
\hat{x}^{\mu }\hat{x}_{\mu }=1/C~~~,~~~\hat{M}_{\mu \nu }=-\hat{M}_{\nu \mu
}~~~,~~~2\hat{x}^{\mu }\hat{M}_{\mu \nu }=\hat{p}_{\nu }/C-A\hat{x}_{\nu }
\label{xhat Mhat conds}
\end{equation}%
We then see that the $M$'s generate $\mathbb{SO}\left( p+1,q\right) $ in the
case of $C>0$ because $sign\left( \eta \right) =\left( p+1,q\right) $.
Similarly the $M$'s generate $\mathbb{SO}\left( p,q+1\right) $ in the case
of $C<0$\ because $sign\left( \eta \right) =\left( p,q+1\right) $. We
expected to see these groups in the group of observables because they are
the symmetry groups for hyperboloids defined by $x^{\mu }x_{\mu }=1/C$.

The enveloping algebra of these operators gives the algebra of observables
on $T^{\ast }M_{C_{p,q}}$ a general element being:%
\begin{equation*}
\hat{f}\left( \hat{x},\hat{M}\right) =\sum_{lm}\tilde{f}_{j,l,m,\mu
_{1}\cdots \mu _{l}}^{\nu _{1}\cdots \nu _{2m}}SYM\left( \hat{x}^{\mu
_{1}}\cdots \hat{x}^{\mu _{l}}\hat{M}_{\nu _{1}\nu _{2}}\cdots \hat{M}_{\nu
_{\left( 2m-1\right) }\nu _{2m}}\right)
\end{equation*}%
where the coefficients $f_{\mu _{1}\cdots \mu _{l}}^{\nu _{1}\cdots \nu
_{2m}}$ are constants.

\subsection{The Algebra of Observables is the Enveloping Algebra of a
Pseudo-Orthogonal Group}

Now that we have a basis of the algebra of observables we want analyze the
Lie group associated to the Lie algebra relations in $\left( \text{%
%TCIMACRO{\TeXButton{\ref{comm xhatMhat}}{\ref{comm xhatMhat}}}%
%BeginExpansion
\ref{comm xhatMhat}%
%EndExpansion
}\right) $. It turns out that the group is $\mathbb{SO}\left( p+1,q+1\right) 
$.

The commutation relations in $\left( \text{%
%TCIMACRO{\TeXButton{\ref{xhat Mhat conds}}{\ref{xhat Mhat conds}}}%
%BeginExpansion
\ref{xhat Mhat conds}%
%EndExpansion
}\right) $ are computed to be equivalent to:%
\begin{equation}
\left[ \hat{M}_{\mu ^{\prime }\nu ^{\prime }},\hat{M}_{\rho ^{\prime }\sigma
^{\prime }}\right] =i\hbar \left( \hat{M}_{\rho ^{\prime }[\mu ^{\prime
}}\eta _{\nu ^{\prime }]\sigma ^{\prime }}-\hat{M}_{\sigma ^{\prime }[\mu
^{\prime }}\eta _{\nu ^{\prime }]\rho ^{\prime }}\right) 
\label{comm Mhatprime}
\end{equation}%
\begin{equation*}
\hat{M}_{\mu ^{\prime }\nu ^{\prime }}=-\hat{M}_{\nu ^{\prime }\mu ^{\prime
}}
\end{equation*}%
where we use the notation that the primed indices run from $1,\ldots ,n+2$.
Thus the $\hat{M}^{\prime }$'s (i.e., the $\hat{M}_{\mu ^{\prime }\nu
^{\prime }}$'s) form the Lie Algebra of $\mathbb{SO}\left( p+1,q+1\right) $, 
$\mathfrak{so}\left( p+1,q+1\right) $ for both $C>0$ and $C<0$!\pagebreak 

The extra $n+1$ generators of $\hat{M}$ being:%
\begin{equation*}
\hat{M}_{\left( n+2\right) \mu ^{\prime }}=-\hat{M}_{\mu ^{\prime }\left(
n+2\right) }=\frac{1}{2\sqrt{\left\vert C\right\vert }}\hat{p}_{\mu ^{\prime
}}=\frac{1}{2\sqrt{\left\vert C\right\vert }}\left( C\hat{x}^{\nu }\hat{M}%
_{\nu \mu ^{\prime }}-\frac{CA}{i\hbar }\hat{x}_{\mu ^{\prime }}\right) 
\text{ for }\mu ^{\prime }=1,\ldots ,n+1
\end{equation*}%
\begin{equation*}
\hat{M}_{\left( n+2\right) \left( n+2\right) }=0
\end{equation*}%
along with the extra components of $\eta $ being: 
\begin{equation*}
\eta _{\left( n+2\right) \left( n+2\right) }=-C/\left\vert C\right\vert
\end{equation*}%
\begin{equation*}
\eta _{\left( n+2\right) \mu ^{\prime }}=0\text{ \ for }\mu ^{\prime }\neq
n+2
\end{equation*}%
It is a straightforward computation to verify that the commutation relation $%
\left[ \hat{M}_{\mu ^{\prime }\nu ^{\prime }},\hat{M}_{\rho ^{\prime }\sigma
^{\prime }}\right] $ is the above.

\noindent \underline{The Summary of the Results:}\noindent

\noindent We now have the following scheme worked out exactly:

\begin{itemize}
\item For the configuration space $M_{C_{p,q}}$ with $sign\left( g\right)
=\left( p,q\right) $\ and \ $C>0$: 
\begin{eqnarray*}
&\implies &sign\left( \eta \right) =\left( p+1,q\right) ~~,~~M\text{
generate }\mathbb{SO}\left( p+1,q\right) \\
&\implies &sign\left( \eta ^{\prime }\right) =\left( p+1,q+1\right)
~~,~~M^{\prime }=\left( M,x\right) \text{ generate }\mathbb{SO}\left(
p+1,q+1\right)
\end{eqnarray*}

\item For the configuration space $M_{C_{p,q}}$ with $sign\left( g\right)
=\left( p,q\right) $\ and \ $C<0$:%
\begin{eqnarray*}
&\implies &sign\left( \eta \right) =\left( p,q+1\right) ~~,~~M\text{
generate }\mathbb{SO}\left( p,q+1\right) \\
&\implies &sign\left( \eta ^{\prime }\right) =\left( p+1,q+1\right)
~~,~~M^{\prime }=\left( M,x\right) \text{ generate }\mathbb{SO}\left(
p+1,q+1\right)
\end{eqnarray*}
\end{itemize}

\subsection{A Summary of Results for de Sitter and Anti-de Sitter Space-Times%
}

Here we give a summary of the results we have obtained for the de Sitter and
Anti-de Sitter (dS/AdS) space-times. In the next subsection we will state
the more general results obtained in this paper which is a straightforward
generalization of this case.

We first embed dS/AdS in a flat five dimensional space given by the
embedding formulas:%
\begin{equation*}
\eta _{\mu \nu }x^{\mu }x^{\nu }=1/C\text{ \ \ and \ \ }x^{\mu }p_{\mu }=A
\end{equation*}

where $C$ and $A$ are some real arbitrary constants, and $\eta $ is the
embedding flat metric. For dS $\eta =diag\left( 1,-1,-1,-1,-1\right) $, $C<0$
and AdS $\eta =diag\left( 1,1,-1,-1,-1\right) $, $C>0$.

We obtained the exact results for the Fedosov star-commutators:%
\begin{equation}
\left[ x^{\mu },x^{\nu }\right] _{\ast }=0~~\ ~~[x_{\mu },M_{\nu \rho
}]_{\ast }=i\hbar x_{[\nu }\eta _{\rho ]\mu }
\end{equation}%
\begin{equation*}
\lbrack M_{\mu \nu },M_{\rho \sigma }]_{\ast }=i\hbar (M_{\rho \lbrack \mu
}\eta _{\nu ]\sigma }-M_{\sigma \lbrack \mu }\eta _{\nu ]\rho })
\end{equation*}%
indices run from $0$ to $4$, $M_{\mu \nu }=x_{[\mu }\ast p_{\nu ]}$, $x_{\mu
}=\eta _{\mu \nu }x^{\nu }$.

The conditions of the embedding $x^{\mu }x_{\mu },~x^{\mu }p_{\mu }$ become
the Casimir invariants of the algebra in group theoretic language.

We now summarize our two key observations:

\begin{enumerate}
\item $M$'s generate $\mathbb{SO}\left( 1,4\right) $ and $\mathbb{SO}\left(
2,3\right) $ for dS and AdS respectively.

\item $M$'s and $x$'s generate $\mathbb{SO}\left( 2,4\right) $ for \textit{%
both} dS and AdS.
\end{enumerate}

By calculating $R=-16C$ and $p_{\mu }\ast p^{\mu }$ in terms of $M$ and $x$
the Hamiltonian $\left( \text{%
%TCIMACRO{\TeXButton{\ref{H}}{\ref{H}}}%
%BeginExpansion
\ref{H}%
%EndExpansion
}\right) $ is:%
\begin{equation}
H=2CM_{\mu \nu }\ast M^{\mu \nu }+\left( A-4i\hbar \right) AC-16\xi C
\end{equation}%
where $M_{\mu \nu }\ast M^{\mu \nu }$ is a Casimir invariant of the subgroup 
$\mathbb{SO}\left( 1,4\right) $ or $\mathbb{SO}\left( 2,3\right) $ for dS or
AdS respectively.

In the more familiar form of Hilbert space language the KG equation $\left( 
\text{%
%TCIMACRO{\TeXButton{\ref{KG}}{\ref{KG}}}%
%BeginExpansion
\ref{KG}%
%EndExpansion
}\right) $ takes the form:%
\begin{equation}
(2C\hat{M}_{\mu \nu }\hat{M}^{\mu \nu }+\chi C)\left\vert \phi
_{m}\right\rangle =m^{2}\left\vert \phi _{m}\right\rangle  \label{KG1}
\end{equation}%
where $\left\langle \phi _{m}|\phi _{m}\right\rangle =1$, $%
%TCIMACRO{\U{2102} }%
%BeginExpansion
\mathbb{C}
%EndExpansion
\ni \chi =\left( A-4i\hbar \right) A-16\xi $ is an arbitrary constant, and
we regard all groups to be in a standard irreducible representation on the
set of linear Hilbert space operators.

These subgroups are the symmetry groups of the manifolds for dS or AdS
respectively. Again, $\hat{M}_{\mu \nu }\hat{M}^{\mu \nu }$ is a Casimir
invariant of the subgroup $\mathbb{SO}\left( 1,4\right) $ or $\mathbb{SO}%
\left( 2,3\right) $ for dS or AdS respectively. Therefore, the above KG
equation $\left( \text{%
%TCIMACRO{\TeXButton{\ref{KG1}}{\ref{KG1}}}%
%BeginExpansion
\ref{KG1}%
%EndExpansion
}\right) $ states that the eigenstates of mass $\left\vert \phi
_{m}\right\rangle $ label the different representations of $\mathbb{SO}%
\left( 1,4\right) $ and $\mathbb{SO}\left( 2,3\right) $ for dS and AdS
respectively sitting inside the full group of observables $\mathbb{SO}\left(
2,4\right) $ which is confirmed by the well-known results of 
%TCIMACRO{%
%\TeXButton{Fr\o nsdal C. (1965, 1973, 1975a, 1975b)}{\hyperlink{ref1}{Fr\o nsdal C. (1965, 1973, 1975a, 1975b)}} }%
%BeginExpansion
\hyperlink{ref1}{Fr\o nsdal C. (1965, 1973, 1975a, 1975b)}
%EndExpansion
as well as others.

\begin{description}
\item[E.g.] In the case of spin 0 particles the operator $\hat{M}^{2}$\
becomes the Laplace-Beltrami operator $\nabla _{\mu }\nabla ^{\mu }$ and $%
\hat{x}^{\mu }\hat{p}_{\mu }\rightarrow -i\hbar x^{\mu }\nabla _{\mu }$ so
let $\phi \left( x\right) :=\left\langle x|\phi \right\rangle $ then:%
\begin{equation*}
\left( 2i\hbar C\nabla _{\mu }\nabla ^{\mu }-\chi C-m^{2}\right) \phi \left(
x\right) =0
\end{equation*}%
where $-i\hbar x^{\mu }\nabla _{\mu }\phi =A\phi $. This equation is the
free wave equation on AdS that is studied in 
%TCIMACRO{%
%\TeXButton{Fr\o nsdal C. (1973)}{\hyperlink{ref1}{Fr\o nsdal C. (1973)}} }%
%BeginExpansion
\hyperlink{ref1}{Fr\o nsdal C. (1973)}
%EndExpansion
and therefore the results given here are consistent with what has been done
previously.
\end{description}

\subsection{The Algebra of Observables and the Klein-Gordon (KG) Equation in
the Our Case}

This subsection is a straightforward generalization of the last subsection.
This summarizes the main results of this paper in its most general form.

We rewrite $p_{\mu }\ast p^{\mu }$ in terms of the generators of all groups
and subgroups (i.e., $x$'s and the $M$'s) and the Casimir invariants of the
these groups and subgroups.

It is well-known that the Casimir invariants of the subgroup generated by $M$
are:%
\begin{eqnarray*}
M^{2} &:&=M_{\mu \nu }\ast M^{\mu \nu } \\
M^{4} &:&=M_{\mu _{1}\mu _{2}}\ast M^{\mu _{2}\mu _{3}}\ast M_{\mu _{3}\mu
_{4}}\ast M^{\mu _{4}\mu _{1}} \\
&&\vdots \\
M^{N} &:&=M_{\mu _{1}\mu _{2}}\ast M^{\mu _{2}\mu _{3}}\ast \cdots \ast
M_{\mu _{N-1}\mu _{N}}\ast M^{\mu _{N}\mu _{1}}
\end{eqnarray*}

where $N$ is the integer part of $\frac{p+q+1}{2}$, i.e., the rank of the
group $\mathbb{SO}\left( p+1,q\right) $ or $\mathbb{SO}\left( p,q+1\right) $.

Also, the Casimir invariants of the full group $\mathbb{SO}\left(
p+1,q+1\right) $ are:%
\begin{eqnarray*}
M^{\prime 2} &:&=M_{\mu ^{\prime }\nu ^{\prime }}\ast M^{\mu ^{\prime }\nu
^{\prime }} \\
M^{\prime 4} &:&=M_{\mu _{1}^{\prime }\mu _{2}^{\prime }}\ast M^{\mu
_{2}^{\prime }\mu _{3}^{\prime }}\ast M_{\mu _{3}^{\prime }\mu _{4}^{\prime
}}\ast M^{\mu _{4}^{\prime }\mu _{1}^{\prime }} \\
&&\vdots \\
M^{\prime N^{\prime }} &:&=M_{\mu _{1}^{\prime }\mu _{2}^{\prime }}\ast
M^{\mu _{2}^{\prime }\mu _{3}^{\prime }}\ast \cdots \ast M_{\mu _{N^{\prime
}-1}^{\prime }\mu _{N^{\prime }}^{\prime }}\ast M^{\mu _{N^{\prime
}}^{\prime }\mu _{1}^{\prime }}
\end{eqnarray*}%
where $N^{\prime }$ is the integer part of $\frac{p+q+2}{2}$, i.e., the rank
of the group $\mathbb{SO}\left( p+1,q+1\right) $.

Using the equation $M_{\mu \nu }=x_{[\mu }\ast p_{\nu ]}$ we compute
directly:%
\begin{equation*}
M^{\prime 2}=-\frac{1}{2}\left( A-i\hbar n\right) A
\end{equation*}%
\begin{equation*}
M^{2}=\frac{1}{2C}p_{\mu }\ast p^{\mu }+M^{\prime 2}\implies p_{\mu }\ast
p^{\mu }=2C\left( M^{2}-M^{\prime 2}\right) 
\end{equation*}%
using $\left[ x^{\mu },p_{\mu }\right] _{\ast }=i\hbar \left( \delta _{\mu
}^{\mu }-1\right) =n$ is the dimension of $M$, $R=-n^{2}C$, $x^{\mu }\ast
x_{\mu }=1/C$, and $x^{\mu }\ast p_{\mu }=A$.\pagebreak 

So by calculating $R=-n^{2}C$ and $p_{\mu }\ast p^{\mu }$ in terms of $M$
and $x$ the Hamiltonian $\left( \text{%
%TCIMACRO{\TeXButton{\ref{H}}{\ref{H}}}%
%BeginExpansion
\ref{H}%
%EndExpansion
}\right) $ is:%
\begin{equation}
H=p_{\mu }\ast p^{\mu }+\xi R=2CM^{2}+C\left( A-i\hbar n\right) A-n^{2}\xi C
\label{H2}
\end{equation}%
where $M_{\mu \nu }\ast M^{\mu \nu }$ is a Casimir invariant of the subgroup 
$\mathbb{SO}\left( p,q+1\right) $ or $\mathbb{SO}\left( p+1,q\right) $ for $%
T^{\ast }M_{C_{p,q}}^{-}$ and $T^{\ast }M_{C_{p,q}}^{+}$respectively. In
addition these subgroups are the symmetry groups of the manifolds for $%
T^{\ast }M_{C_{p,q}}^{-}$ and $T^{\ast }M_{C_{p,q}}^{+}$ respectively.

Using the correspondence between a Hilbert space formulation and DQ given by
Fedosov as mentioned in the last section we reformulate $\left( \text{%
%TCIMACRO{\TeXButton{\ref{KGDQ}}{\ref{KGDQ}}}%
%BeginExpansion
\ref{KGDQ}%
%EndExpansion
}\right) $ into the form of $\left( \text{%
%TCIMACRO{\TeXButton{\ref{KG}}{\ref{KG}}}%
%BeginExpansion
\ref{KG}%
%EndExpansion
}\right) $.

In the more familiar form of Hilbert space language the KG equation $\left( 
\text{%
%TCIMACRO{\TeXButton{\ref{KG}}{\ref{KG}}}%
%BeginExpansion
\ref{KG}%
%EndExpansion
}\right) $ takes the form:%
\begin{equation}
(2C\hat{M}_{\mu \nu }\hat{M}^{\mu \nu }+\chi C)\left\vert \phi
_{m}\right\rangle =m^{2}\left\vert \phi _{m}\right\rangle  \label{KG2}
\end{equation}%
where $\left\langle \phi _{m}|\phi _{m}\right\rangle =1$, $%
%TCIMACRO{\U{2102} }%
%BeginExpansion
\mathbb{C}
%EndExpansion
\ni \chi =\left( A-i\hbar n\right) A-n^{2}\xi $ is an arbitrary constant,
and we regard all groups to be in a standard irreducible representation on
the set of linear Hilbert space operators. Again, $\hat{M}_{\mu \nu }\hat{M}%
^{\mu \nu }$ is a Casimir invariant of the subgroup $\mathbb{SO}\left(
p,q+1\right) $ or $\mathbb{SO}\left( p+1,q\right) $ for $T^{\ast
}M_{C_{p,q}}^{-}$ or $T^{\ast }M_{C_{p,q}}^{+}$ respectively.

Therefore, the above KG equation $\left( \text{%
%TCIMACRO{\TeXButton{\ref{KG2}}{\ref{KG2}}}%
%BeginExpansion
\ref{KG2}%
%EndExpansion
}\right) $ states that the eigenstates of mass $\left\vert \phi
_{m}\right\rangle $ label the different representations of $\mathbb{SO}%
\left( p,q+1\right) $ and $\mathbb{SO}\left( p+1,q\right) $ for dS and AdS
respectively sitting inside the full group of observables $\mathbb{SO}\left(
p+1,q+1\right) $.

\begin{description}
\item[E.g.] In the case of spin 0 particles in $n$-dimensions the operator $%
\hat{M}^{2}$\ becomes the Laplace-Beltrami operator $\nabla _{\mu }\nabla
^{\mu }=\left( -g\right) ^{1/2}\partial _{\mu }g^{\mu \nu }\left( -g\right)
^{-1/2}\partial _{\nu }$ and $\hat{x}^{\mu }\hat{p}_{\mu }\rightarrow
-i\hbar x^{\mu }\nabla _{\mu }$ so let $\phi \left( x\right) :=\left\langle
x|\phi \right\rangle $ then:%
\begin{equation*}
\left( 2C\nabla _{\mu }\nabla ^{\mu }-C\chi -m^{2}\right) \phi \left(
x\right) =0
\end{equation*}%
where $x^{\mu }\nabla _{\mu }\phi =A\phi $. This equation is the free scalar
wave equation on $T^{\ast }M_{C_{p,q}}^{\pm }$.
\end{description}

\section{Conclusions}

In conclusion, the results of this paper confirm the well known results for
the Klein-Gordon equation in 
%TCIMACRO{%
%\TeXButton{Fr\o nsdal C. (1965, 1973, 1975a, 1975b)}{\hyperlink{ref1}{Fr\o nsdal C. (1965, 1973, 1975a, 1975b)}} }%
%BeginExpansion
\hyperlink{ref1}{Fr\o nsdal C. (1965, 1973, 1975a, 1975b)}
%EndExpansion
as well as many others. The difference is that we confirmed these results in
the context of DQ. The beautiful thing about these computations is that they
are algorithmic and they can be done for any manifold, whereas some previous
techniques in quantization relied heavily on the symmetries of these
particular manifolds or the type of dynamical evolutions studied. We note
that while we expected the symmetry group of the observables $\mathbb{SO}%
\left( q,p+1\right) $ or $\mathbb{SO}\left( q+1,p\right) $ to be in this
group we did not expect that the full group of observables to be $\mathbb{SO}%
\left( q+1,p+1\right) $. This fact may be well-known to group theorists,
however it was surprising to us. In the dS/AdS this is the group $\mathbb{SO}%
\left( 2,4\right) $ this we suspect is the conformal group of the manifold $%
\mathbb{SO}\left( 2,4\right) $ but a clear interpretation is needed to
assert this claim.

\section{Acknowledgements}

We would like to thank E. Ted Newman and Al Janis for their helpful
comments. Also we would like to thank the Laboratory of
Axiomatics.\pagebreak 

\section{Appendix A: Notations}

%TCIMACRO{%
%\TeXButton{\hypertarget{Appendix A}{}}{\hypertarget{Appendix A}{}}}%
%BeginExpansion
\hypertarget{Appendix A}{}%
%EndExpansion
Here I will briefly list mention my definitions and notations:

\begin{enumerate}
\item \textbf{Index Notations:}

\begin{enumerate}
\item We use the convention that the lower case indices run from $1,\ldots
,n $ (space-time indices)\ and capital ones run from $1,\ldots 2n$
(phase-space indices).

\item We employ the abstract index notation for this paper for lowercase
indices only. Lower-case greek letters are numerical indices while
lower-case latin letters are abstract ones. (See 
%TCIMACRO{\TeXButton{Wald R. 1984}{\hyperlink{ref1}{Wald R. 1984}}}%
%BeginExpansion
\hyperlink{ref1}{Wald R. 1984}%
%EndExpansion
).

\item The abstract indices that are not written will be form indices so that
multiplication of them implies a wedging $\wedge $ of the forms.

\textbf{Abstract indices convention:}

When we write $D=\Theta ^{B}D_{B}$ and this acts on some configuration space
quantity like a one-form $v_{a}=v_{\mu }\left( x\right) dx^{\mu }$ (on the
configuration space) in the operator $D$ the tensor index is suppressed. We
therefore make the convention that in the abstract index notation the label $%
B$ in $D=\Theta ^{B}D_{B}$ will determine the abstract index of the
configuration space quantity as $b$. For example:%
\begin{equation*}
D\otimes v_{a}=\Theta ^{B}D_{B}\otimes v_{a}=\nabla _{b}v_{a}
\end{equation*}

\item Some exceptions to our index convention is needed. The letters $%
j,l,m,k $ will \textbf{always}\ be reserved for labelling powers and other
numerical labelling including non-space-time indices and thus will not go
according to our index conventions in a. and b.
\end{enumerate}

\item \textbf{Raising and lowering indices:} We will always raise and lower
the lower-case indices or $M_{C_{p,q}}$\ indices (greek or latin) by the
metric of the imbedding space $\eta _{\mu \nu }$. We will always raise and
lower the upper-case indices with the symplectic form $\omega _{AB}$.

\item \textbf{Constant curvature manifold of codimension one:}

An $n$ dimensional constant curvature manifold embedded in a $\left(
n+1\right) $-dimensional flat space ($%
%TCIMACRO{\U{211d} }%
%BeginExpansion
\mathbb{R}
%EndExpansion
^{n+1}$) given by an embedding 
\begin{equation*}
\eta _{\mu \nu }x^{\mu }x^{\nu }=1/C~~~,~~~x^{\mu }p_{\mu }=A
\end{equation*}%
where $\mu =1,\ldots ,n+1$.%
\begin{equation*}
T^{\ast }M_{C_{p,q}}^{+}=T^{\ast }M_{C_{p,q}}\text{ with }%
C>0~~,~~~sign\left( g\right) =\left( p,q\right) ~~~,~~~sign\left( \eta
\right) =\left( p+1,q\right)
\end{equation*}%
\begin{equation*}
T^{\ast }M_{C_{p,q}}^{-}=T^{\ast }M_{C_{p,q}}\text{ with }%
C<0~~,~~~sign\left( g\right) =\left( p,q\right) ~~~,~~~sign\left( \eta
\right) =\left( p,q+1\right)
\end{equation*}%
where $sign\left( g\right) $ is the signature of the metric i.e.%
\begin{equation*}
sign\left( g\right) =\left( p,q\right) \implies g=-\left( d\tilde{x}%
^{1}\right) ^{2}-\ldots -\left( d\tilde{x}^{q}\right) ^{2}+\left( d\tilde{x}%
^{q+1}\right) ^{2}+\ldots +\left( d\tilde{x}^{p+q}\right) ^{2}
\end{equation*}%
in some local coordinates $\tilde{x}^{\mu }$. For example $T^{\ast
}M_{C_{1,3}}^{-}$ is dS and $T^{\ast }M_{C_{1,3}}^{+}$ is AdS.

\item \textbf{Configuration space connection and curvature on the constant
curvature manifold of codimension one case (}$M_{C_{p,q}}$\textbf{): }We let 
$\partial _{a}$ be the flat embedding connection of the ambient space and $%
\nabla _{a}$ to be the connection on the manifold $M_{C_{p,q}}$. Let $f$ be
an arbitrary function and let $dx^{\mu }$ be a basis of forms on the
manifold $M_{C_{p,q}}$ then:%
\begin{equation*}
\nabla _{\sigma }f\left( x\right) =\frac{\partial f}{\partial x^{\sigma }}
\end{equation*}%
\begin{equation*}
\nabla _{\sigma }\left( dx^{\mu }\right) =-\Gamma _{~\nu \sigma }^{\mu
}dx^{\nu }
\end{equation*}%
\begin{equation*}
\nabla _{\sigma }\left( \frac{\partial }{\partial x^{\mu }}\right) =\Gamma
_{~\mu \sigma }^{\nu }\frac{\partial }{\partial x^{\nu }}
\end{equation*}%
\begin{equation*}
\nabla _{\lbrack \sigma }\nabla _{\rho ]}\left( dx^{\mu }\right) =R_{~\nu
\sigma \rho }^{\mu }dx^{\nu }
\end{equation*}%
We can extend to higher order tensors by using the Leibnitz rule and the
fact that $\nabla $ commutes with contractions.

\item \textbf{Symmetrization and anti-symmetrization of indices:}%
\begin{equation*}
2\Delta _{~\mu \nu }=\Delta _{~\mu \nu }-\Delta _{~\nu \mu }
\end{equation*}%
\begin{equation*}
2\Delta _{~\left( \mu \nu \right) }=\Delta _{~\mu \nu }+\Delta _{~\nu \mu }
\end{equation*}%
\begin{equation*}
n!\Delta _{~\left( \mu _{1}\cdots \mu _{n}\right) }=\Delta _{~\mu _{1}\cdots
\mu _{n}}+\left( \text{all perms}\right)
\end{equation*}%
\begin{equation*}
n!\Delta _{~\left[ \mu _{1}\cdots \mu _{n}\right] }=\Delta _{~\mu _{1}\cdots
\mu _{n}}+\left( \text{even perms}\right) -\left( \text{odd perms}\right)
\end{equation*}

\item \textbf{Coordinates }and the corresponding \textbf{basis of one-forms}
on phase-space where$\left\{ q^{A}\right\} =\left( q^{1},\ldots
,q^{2n}\right) $:%
\begin{equation*}
\left\{ dq^{1},\ldots ,dq^{2n}\right\}
\end{equation*}

\item \textbf{Phase-Space Connection: }Given an arbitrary function $f$ and
vector $v^{B}$\ on phase-space, a\textbf{\ general (torsion-free)
phase-space connection:}%
\begin{equation*}
D_{A}f~=\frac{\partial f}{\partial q^{A}}~~~,~~~D_{A}v^{B}=-\Gamma
_{~BC}^{A}v^{C}
\end{equation*}%
with the conditions that $D$ preserves the symplectic form $D\otimes \omega
=0$ and is torsion-free $D^{2}f=0$ (or in abstract indices: $D_{A}\omega
_{BC}=0$ and $D_{[A}D_{B]}f=0$). We define the connection in the coordinates 
$q^{A}$ is: 
\begin{equation*}
D\otimes \Theta ^{A}=\Gamma _{~C}^{A}\otimes \Theta ^{C}=\Gamma
_{~CB}^{A}\Theta ^{B}\otimes \Theta ^{C}
\end{equation*}%
and the curvature is:%
\begin{equation*}
D^{2}\otimes \Theta ^{A}:=R_{B}^{\text{ \ }A}\otimes \Theta ^{B}=R_{CEB}^{%
\text{ \ ~~~~}A}\Theta ^{C}\wedge \Theta ^{E}\otimes \Theta ^{B}
\end{equation*}%
*We note that these conditions do not specify $D_{A}$ uniquely. We are free
to add a tensor $\Delta _{ABC}$ symmetric in $\left( ABC\right) $, i.e., a
new connection $D_{new}$ may be defined by:%
\begin{equation*}
D_{new}\otimes \Theta ^{A}=\Gamma _{~CB}^{A}\Theta ^{B}\otimes \Theta
^{C}+\Delta _{~CB}^{A}\Theta ^{B}\otimes \Theta ^{C}
\end{equation*}%
Again, we can extend to higher order tensors by using the Leibnitz rule and
the fact that $D$ commutes with contractions.

\item \textbf{Flat connection:} When the phase-space is flat, i.e.,
associated to a flat space/space-time we will use $\partial _{A}$ instead of 
$D_{A}$ for the connection.

\item \textbf{Antisymmetric and symmetric tensor products:} The wedge
product $\wedge $ is reserved for the antisymmetric tensor product%
\begin{equation*}
\theta \wedge \alpha :=\theta \otimes \alpha -\alpha \otimes \theta
\end{equation*}%
and the vee product $\vee $ is reserved for the symmetric tensor product:%
\begin{equation*}
\theta \vee \alpha :=\theta \otimes \alpha +\alpha \otimes \theta
\end{equation*}%
Since writing $\wedge $ and $\vee $ all over the place will become
cumbersome we will make the convention that we will not write them because
it will be obvious when we mean one or the other. For example, the metric
always uses the symmetric tensor product $g=g_{\mu \nu }dx^{\mu }\vee
dx^{\nu }$ and the symplectic form always uses the antisymmetric tensor
product $\omega =\omega _{AB}\Theta ^{A}\wedge \Theta ^{B}$. However, we
simply write them $g=g_{\mu \nu }dx^{\mu }dx^{\nu }$ and $\omega =\omega
_{AB}\Theta ^{A}\Theta ^{B}$.

\item Also, when we write $D^{2}$ or $(D-\hat{D})^{2}$ like in equations $%
\left( \text{%
%TCIMACRO{\TeXButton{\ref{cond_Dhat}}{\ref{cond_Dhat}}}%
%BeginExpansion
\ref{cond_Dhat}%
%EndExpansion
}\right) $ and $\left( \text{%
%TCIMACRO{\TeXButton{\ref{D2}}{\ref{D2}}}%
%BeginExpansion
\ref{D2}%
%EndExpansion
}\right) $ we always mean antisymmetric tensor products because these are
curvature equations. In the curvature operators like $D^{2}$ or $(D-\hat{D}%
)^{2}$ the $\Theta $'s are always wedged together by definition. An example
is:%
\begin{equation*}
D^{2}v_{B}=R_{B}^{\text{ \ }A}v_{A}=R_{CEB}^{\text{ \ ~~~}A}\Theta
^{C}\wedge \Theta ^{E}v_{A}
\end{equation*}%
If indices $A$, $B$, $C,$ etc. are all abstract then the formula above is:%
\begin{equation*}
\left[ D_{A},D_{B}\right] v_{B}=2R_{CEB}^{\text{ \ ~~~}A}v_{A}
\end{equation*}

\item \textbf{The symplectic form:}%
\begin{equation*}
\omega =\omega _{AB}\Theta ^{A}\wedge \Theta ^{B}=\omega _{AB}\Theta
^{A}\Theta ^{B}
\end{equation*}%
$\omega ^{AB}$ is the inverse of $\omega _{AB}$\ with $\omega ^{AB}\omega
_{BC}=\delta _{C}^{A}$.

\item \textbf{The Poisson bracket:}%
\begin{equation*}
\left[ f,g\right] _{P}=\omega ^{AB}\left( D_{A}f\right) \left( D_{B}g\right)
\end{equation*}%
\begin{equation*}
\overleftrightarrow{P}:=\overleftarrow{D}_{A}\omega ^{AB}\overrightarrow{D}%
_{B}~~~,~~~f\overleftrightarrow{P}g=\left[ f,g\right] _{P}
\end{equation*}%
where $f$ and $g$ are two arbitrary functions and the arrows determine the
direction that each derivative acts it.

\item \textbf{Darboux coordinates and Darboux's Theorem:} In the
neighborhood of each point on an $n$-dimensional symplectic manifold, there
exists coordinates called \textbf{Darboux coordinates} $\tilde{q}=\left( 
\tilde{x}^{1},\ldots ,\tilde{x}^{n},\tilde{p}_{1},\ldots ,\tilde{p}%
_{n}\right) $\footnote{%
Note that these $2n$ coordinates and are different from the $2n+2$ embedding
coordinates $\left( x^{\mu },p_{\mu }\right) $.} where the $\omega $ takes
the form:%
\begin{equation*}
\omega =d\tilde{p}_{1}d\tilde{x}^{1}+\cdots +d\tilde{p}_{n}d\tilde{x}^{n}
\end{equation*}

\item \textbf{Groenewold-Moyal star:} In terms of the flat connection $%
\partial $ Groenewold-Moyal star is:%
\begin{equation}
f\ast g=fe^{\frac{i\hbar }{2}\overleftarrow{\partial }_{A}\omega ^{AB}%
\overrightarrow{\partial }_{B}}g=fg+\frac{i\hbar }{2}\omega ^{AB}\left(
\partial _{A}f\right) \left( \partial _{B}g\right) -\frac{\hbar ^{2}}{8}%
\omega ^{CE}\omega ^{AB}\left( \partial _{C}\partial _{A}f\right) \left(
\partial _{E}\partial _{B}g\right) +\cdots  \label{Moyal1}
\end{equation}%
\begin{equation*}
f\ast g=\sum_{A,B,j}^{\infty }\left( i\hbar /2\right) ^{j}\omega
^{A_{1}B_{1}}\cdots \omega ^{A_{j}B_{j}}/j!(\partial _{A_{1}}\cdots \partial
_{A_{j}}f)(\partial _{B_{1}}\cdots \partial _{B_{j}}g)
\end{equation*}

\item Smooth functions on a space $A$, $C^{\infty }\left( A\right) $.

\item The \textbf{traces over translational degrees of freedom:}%
\begin{equation*}
Tr_{tr}(\hat{f}):=\int d^{n}x~\langle x|\hat{f}|x\rangle
\end{equation*}%
\begin{equation*}
Tr_{tr\ast }\left( f\right) :=\frac{1}{\left( 2\pi \hbar \right) ^{n}}\int
d^{n}pd^{n}x~f=\frac{1}{\left( 2\pi \hbar \right) ^{n}}\int d^{2n}q~f
\end{equation*}

\item The \textbf{traces over all degrees of freedom}, i.e., over the
translation degrees of freedom as well as all other degrees of freedom is
denoted by $Tr$ and $Tr_{\ast }$.

\item Let $\left( N,\omega \right) $ be a \textbf{symplectic manifold} where 
$\omega $ is a nondegenerate closed ($d\omega =0$) two-form.

\item \textbf{Formal series in} $\hbar $ is a power series in $\hbar $ with
coefficients in $A$ denoted by adding $\left[ \left[ \hbar \right] \right] $
like $A\left[ \left[ \hbar \right] \right] $.

For example, $C^{\infty }\left( T^{\ast }M\right) \left[ \left[ \hbar \right]
\right] $ is formal series in $\hbar $ with coefficients in $C^{\infty
}\left( T^{\ast }M\right) $. Let $f\left( q\right) \in C^{\infty }\left(
T^{\ast }M\right) \left[ \left[ \hbar \right] \right] $ then%
\begin{equation*}
f\left( q\right) =f_{j}\left( q\right) ~\hbar ^{j}=f_{0}\left( q\right)
+f_{1}\left( q\right) ~\hbar +f_{2}\left( q\right) ~\hbar ^{2}+\cdots
\end{equation*}%
where $f_{j}\left( q\right) \in C^{\infty }\left( T^{\ast }M\right) $ for
each $j$.

\item \textbf{Star-exponential:}%
\begin{equation*}
\exp _{\ast }\left( f\right) :=e_{\ast }^{f}:=\sum_{j}\left( f\ast \right)
^{j}/j!=1+f+\frac{1}{2!}f\ast f+\frac{1}{3!}f\ast f\ast f+\cdots
\end{equation*}

\item \textbf{Complex-valued and Matrix-valued functions: }In this paper
when we say $f$ is a function/form we define it to be a complex Taylor
series in its variables\footnote{%
The set of all of these type of functions is sometimes called the enveloping
algebra of its arguments.}. Explicitly:%
\begin{equation}
f\left( u,\ldots ,v\right) =\sum_{l,j}f_{j_{1}\cdots j_{l}}u^{j_{1}}\cdots
v^{j_{l}}\text{ \ \ (}j\text{'s are powers not indices)}  \notag
\end{equation}%
where $u$ and $v$ are arbitrary.

So if $f$ is a function/form of some subset or all of the quantities $%
x,p,dx,dp,\omega ,\hbar $ and $i$ it then commutes with the $\hat{y}$'s and
will be called a complex-valued function/form. On the contrary an
matrix-valued function/form is a complex Taylor series in $\hat{y}$ and
possibly some subset or all of the quantities $x,p,dx,dp,\omega ,\hbar $ and 
$i$.

So if $f\left( x,p,dx,dp,\omega ,\hbar ,i\right) $ is a complex-valued
function/form it then commutes with the $\hat{y}$'s. More explicitly with
the matrix indices written (which are exceptions to our index conventions):%
\begin{equation*}
\left( \hat{y}^{A}\hat{y}^{B}\right) _{jk}=\sum_{l}\hat{y}_{jl}^{A}\hat{y}%
_{lk}^{B}
\end{equation*}%
\begin{equation*}
\left( \left[ \hat{y}^{A},f\right] \right) _{jk}:=\hat{y}_{jk}^{A}f-f\hat{y}%
_{jk}^{A}=0
\end{equation*}%
On the contrary a matrix-valued function/form does not. For this paper, we
will not write these matrix indices explicitly.
\end{enumerate}

\section{Appendix B: The Proof of the Integrability of $\left( \text{%
%TCIMACRO{\TeXButton{\ref{cond Qhat}}{\ref{cond Qhat}}}%
%BeginExpansion
\ref{cond Qhat}%
%EndExpansion
}\right) $}

\noindent 
%TCIMACRO{%
%\TeXButton{\hypertarget{Appendix B}{}}{\hypertarget{Appendix B}{}}}%
%BeginExpansion
\hypertarget{Appendix B}{}%
%EndExpansion
We want to show that the condition $\left( \text{%
%TCIMACRO{\TeXButton{\ref{cond Qhat}}{\ref{cond Qhat}}}%
%BeginExpansion
\ref{cond Qhat}%
%EndExpansion
}\right) $ is integrable locally. Showing that the following the $P$ of the
condition in $\left( \text{%
%TCIMACRO{\TeXButton{\ref{cond Qhat}}{\ref{cond Qhat}}}%
%BeginExpansion
\ref{cond Qhat}%
%EndExpansion
}\right) $ vanishes:%
\begin{equation}
P\left( \text{%
%TCIMACRO{\TeXButton{\ref{cond Qhat}}{\ref{cond Qhat}}}%
%BeginExpansion
\ref{cond Qhat}%
%EndExpansion
}\right) =P\left( \left( P-dx^{\mu }\hat{\partial}_{\mu }\right) f_{~\sigma
}^{\nu }+\Gamma _{~\rho \mu }^{\nu }dx^{\mu }f_{~\sigma }^{\rho }-\Gamma
_{~\sigma \mu }^{\nu }dx^{\mu }+R_{~\mu \beta \sigma }^{\nu }s^{\mu
}dx^{\beta }\right) dx^{\sigma }=0  \label{cond Qhat3}
\end{equation}%
where $P$ is the differential operator:%
\begin{equation*}
P=\left( D+f_{~\rho }^{\sigma }dx^{\rho }\hat{\partial}_{\sigma }\right)
\end{equation*}%
where $\hat{\partial}_{\mu }:=\partial /\partial s^{\mu }$ implies that the
condition $\left( \text{%
%TCIMACRO{\TeXButton{\ref{cond Qhat}}{\ref{cond Qhat}}}%
%BeginExpansion
\ref{cond Qhat}%
%EndExpansion
}\right) $ is integrable locally by the Cauchy-Kovalevskaya theorem.

\noindent *Note this analogous to how Fedosov can locally integrate the
solution for $\hat{D}$, i.e., by requiring that $\left( D-\hat{D}\right) ^{2}%
\hat{y}^{A}=0$ in $\left( \text{%
%TCIMACRO{\TeXButton{\ref{cond_Dhat}}{\ref{cond_Dhat}}}%
%BeginExpansion
\ref{cond_Dhat}%
%EndExpansion
}\right) $. However, before doing this by brute force we notice that $D$
acting on everything in the equation above is just the configuration space
connection $\nabla $. Therefore, to simplify the calculation we will us
abstract indices. The equation above in $\left( \text{%
%TCIMACRO{\TeXButton{\ref{cond Qhat3}}{\ref{cond Qhat3}}}%
%BeginExpansion
\ref{cond Qhat3}%
%EndExpansion
}\right) $ (and in $\left( \text{%
%TCIMACRO{\TeXButton{\ref{cond Qhat}}{\ref{cond Qhat}}}%
%BeginExpansion
\ref{cond Qhat}%
%EndExpansion
}\right) $) becomes the equation:%
\begin{equation}
\left( \nabla _{\lbrack n}+f_{~[n}^{d}\hat{\partial}_{d}\right) \left(
R_{ca]~m}^{~~b}s^{m}+\left( \nabla _{c}+f_{~c}^{e}\hat{\partial}%
_{|e|}\right) f_{~a]}^{b}\right) =0  \label{cond Qhat4}
\end{equation}%
where $P$ on configuration space quantities is $\left( \nabla _{n}+f_{~n}^{d}%
\hat{\partial}_{d}\right) $.

First we note that we want $f_{~b}^{a}$ to be a globally defined object
hence it should be made out of tensors. This rules out the trivial solution
of $f_{~\rho }^{\sigma }dx^{\rho }=-\Gamma _{~\rho \nu }^{\sigma }s^{\nu
}dx^{\rho }$.

\noindent \textbf{Proof:}%
\begin{eqnarray*}
\left( \text{%
%TCIMACRO{\TeXButton{\ref{cond Qhat4}}{\ref{cond Qhat4}}}%
%BeginExpansion
\ref{cond Qhat4}%
%EndExpansion
}\right) &=&\left( \nabla _{\lbrack n}+f_{~[n}^{d}\hat{\partial}_{d}\right)
\left( R_{ca]~m}^{~~b}s^{m}+\left( \nabla _{c}+f_{~c}^{e}\hat{\partial}%
_{|e|}\right) f_{~a]}^{b}\right) \\
&=&\left( \nabla _{\lbrack n}R_{ca]~m}^{~~b}\right)
s^{m}-R_{~m[ac}^{b}\left( \nabla _{n]}+f_{~n]}^{d}\hat{\partial}_{d}\right)
s^{m}+\left( \nabla _{\lbrack n}+f_{~[n}^{d}\hat{\partial}_{d}\right) \left(
\nabla _{c}+f_{~c}^{e}\hat{\partial}_{|e|}\right) f_{~a]}^{b}
\end{eqnarray*}%
In abstract indices we have the identities:%
\begin{equation*}
Ds^{a}=0
\end{equation*}%
\begin{equation*}
\nabla _{a}f\left( x,s\right) =\partial _{a}f-\Gamma _{~ab}^{c}s^{b}\hat{%
\partial}_{c}f
\end{equation*}%
\begin{equation*}
\nabla _{\lbrack n}\nabla _{c]}f\left( x,s\right) =R_{~enc}^{b}s^{e}\hat{%
\partial}_{b}f
\end{equation*}%
and we have the second Bianchi identity:%
\begin{equation*}
\nabla _{\lbrack n}R_{ca]~m}^{~~b}=0
\end{equation*}%
We also have the identity:%
\begin{equation}
P^{2}=D^{2}+\frac{4}{3}p_{\delta }R_{~(\beta \kappa )\gamma }^{\delta
}f_{~\rho }^{\beta }dx^{\kappa }dx^{\rho }\left[ s^{\gamma },\cdot \right]
+\left( Pf_{~\rho }^{\sigma }+\Gamma _{~\beta \kappa }^{\sigma }dx^{\kappa
}f_{~\rho }^{\beta }\right) dx^{\rho }\hat{\partial}_{\sigma }  \label{P2_1}
\end{equation}%
\textbf{Proof of }$\left( \text{%
%TCIMACRO{\TeXButton{\ref{P2_1}}{\ref{P2_1}}}%
%BeginExpansion
\ref{P2_1}%
%EndExpansion
}\right) $\textbf{:}%
\begin{eqnarray}
P^{2}h &=&\left( D+f_{~\rho }^{\sigma }dx^{\rho }\hat{\partial}_{\sigma
}\right) ^{2}h  \label{P2_1A} \\
&=&\left( D+f_{~\rho }^{\sigma }dx^{\rho }\hat{\partial}_{\sigma }\right)
\left( D+f_{~\kappa }^{\psi }dx^{\kappa }\hat{\partial}_{\psi }\right) h 
\notag \\
&=&D^{2}h+\underset{\left( C\right) }{\underbrace{f_{~\rho }^{\sigma
}dx^{\rho }\hat{\partial}_{\sigma }\left( Dh\right) +D\left( f_{~\kappa
}^{\psi }dx^{\kappa }\hat{\partial}_{\psi }h\right) }}+\underset{\left(
E\right) }{\underbrace{f_{~\rho }^{\sigma }dx^{\rho }\hat{\partial}_{\sigma
}\left( f_{~\kappa }^{\psi }dx^{\kappa }\hat{\partial}_{\psi }h\right) }} 
\notag
\end{eqnarray}%
We know that:%
\begin{equation*}
\hat{\partial}_{\psi }h=\left[ k_{\psi },h\right] /i\hbar
\end{equation*}%
\begin{equation*}
Dk_{\sigma }:=-\frac{4}{3}R_{~(\sigma \kappa )\beta }^{\delta }dx^{\kappa
}s^{\beta }p_{\delta }+\Gamma _{~\sigma \kappa }^{\rho }dx^{\kappa }k_{\rho }
\end{equation*}%
We can easily prove that $D\left( \left[ f,h\right] \right) -\left[ f,Dh%
\right] =\left[ Df,h\right] $ for any matrix-valued functions $f\left(
x,p,s,k\right) $ and $h\left( x,p,s,k\right) $ therefore:%
\begin{equation*}
i\hbar D\left( \hat{\partial}_{\psi }h\right) -i\hbar \hat{\partial}_{\psi
}\left( Dh\right) =i\hbar \left[ -\frac{4}{3}R_{~(\psi \sigma )\beta
}^{\delta }dx^{\sigma }s^{\beta }p_{\delta }+\Gamma _{~\psi \sigma }^{\rho
}dx^{\sigma }k_{\rho },h\right]
\end{equation*}%
and $\left( C\right) $ becomes:%
\begin{eqnarray*}
\left( C\right) &=&i\hbar \left( f_{~\rho }^{\sigma }dx^{\rho }\hat{\partial}%
_{\sigma }\left( Dh\right) +D\left( f_{~\rho }^{\sigma }dx^{\rho }\hat{%
\partial}_{\sigma }h\right) \right) \\
&=&f_{~\rho }^{\sigma }dx^{\rho }\left( \left[ k_{\sigma },Dh\right]
-D\left( \left[ k_{\sigma },h\right] \right) \right) +D\left( f_{~\rho
}^{\sigma }dx^{\rho }\right) \left[ k_{\sigma },h\right] \\
&=&f_{~\rho }^{\sigma }dx^{\rho }\left[ Dk_{\sigma },h\right] +D\left(
f_{~\rho }^{\sigma }dx^{\rho }\right) \left[ k_{\sigma },h\right] \\
&=&f_{~\rho }^{\psi }dx^{\rho }\left( -\frac{4}{3}R_{~(\psi \kappa )\beta
}^{\delta }dx^{\kappa }\left[ s^{\beta },h\right] p_{\delta }+\Gamma _{~\psi
\kappa }^{\sigma }dx^{\kappa }\left[ k_{\sigma },h\right] \right) +\left(
Df_{~\rho }^{\sigma }\right) dx^{\rho }\left[ k_{\sigma },h\right]
\end{eqnarray*}%
\begin{equation*}
\implies \left( C\right) =\frac{4}{3}p_{\delta }R_{~(\psi \kappa )\beta
}^{\delta }f_{~\rho }^{\psi }dx^{\kappa }dx^{\rho }\left[ s^{\beta },h\right]
+\left( Df_{~\rho }^{\sigma }+f_{~\rho }^{\psi }\Gamma _{~\psi \kappa
}^{\sigma }dx^{\kappa }\right) dx^{\rho }\hat{\partial}_{\sigma }h
\end{equation*}%
also:%
\begin{eqnarray*}
\left( E\right) &=&f_{~\rho }^{\sigma }dx^{\rho }dx^{\kappa }\left( \hat{%
\partial}_{\sigma }f_{~\kappa }^{\psi }\right) \hat{\partial}_{\psi
}h+f_{~\rho }^{\sigma }dx^{\rho }f_{~\kappa }^{\psi }dx^{\kappa }\left( \hat{%
\partial}_{\sigma }\hat{\partial}_{\psi }h\right) \\
&=&f_{~\rho }^{\sigma }dx^{\rho }dx^{\kappa }\left( \hat{\partial}_{\sigma
}f_{~\kappa }^{\psi }\right) \hat{\partial}_{\psi }h
\end{eqnarray*}%
Putting $\left( C\right) $ and $\left( E\right) $ into the condition at $%
\left( \text{%
%TCIMACRO{\TeXButton{\ref{P2_1A}}{\ref{P2_1A}}}%
%BeginExpansion
\ref{P2_1A}%
%EndExpansion
}\right) $:%
\begin{equation*}
P^{2}h=D^{2}h+\frac{4}{3}p_{\delta }R_{~(\psi \kappa )\beta }^{\delta
}f_{~\rho }^{\psi }dx^{\kappa }dx^{\rho }\left[ s^{\beta },h\right] +\left(
Pf_{~\rho }^{\sigma }+f_{~\rho }^{\psi }dx^{\kappa }\Gamma _{~\psi \kappa
}^{\sigma }\right) dx^{\rho }\hat{\partial}_{\sigma }h
\end{equation*}%
\begin{equation*}
\implies P^{2}=D^{2}+\frac{4}{3}p_{\delta }R_{~(\beta \kappa )\gamma
}^{\delta }f_{~\rho }^{\beta }dx^{\kappa }dx^{\rho }\left[ s^{\gamma },\cdot %
\right] +\left( Pf_{~\rho }^{\sigma }+\Gamma _{~\beta \kappa }^{\sigma
}dx^{\kappa }f_{~\rho }^{\beta }\right) dx^{\rho }\hat{\partial}_{\sigma }
\end{equation*}%
\textbf{QED.}

\noindent Using the identity in $\left( \text{%
%TCIMACRO{\TeXButton{\ref{P2_1}}{\ref{P2_1}}}%
%BeginExpansion
\ref{P2_1}%
%EndExpansion
}\right) $ in abstract indices is:%
\begin{eqnarray*}
&&\left( \nabla _{\lbrack n}+f_{~[n}^{d}\hat{\partial}_{d}\right) \left(
\nabla _{c]}+f_{~c]}^{e}\hat{\partial}_{e}\right) h \\
&=&\nabla _{\lbrack n}\nabla _{c]}h+\frac{2}{3}p_{d}\left(
R_{~(mc)a}^{d}f_{~n}^{m}-R_{~(mn)a}^{d}f_{~c}^{m}\right) \left[ s^{a},h%
\right] +\left( \left( \nabla _{\lbrack n}+f_{~[n}^{d}\hat{\partial}%
_{|d|}\right) f_{~c]}^{e}\right) \hat{\partial}_{e}h \\
&=&\nabla _{\lbrack n}\nabla _{c]}h+\frac{2}{3}p_{d}\left(
R_{~(mc)a}^{d}f_{~n}^{m}-R_{~(mn)a}^{d}f_{~c}^{m}\right) \left[ s^{a},h%
\right] +\left( \left( \nabla _{\lbrack n}f_{~c]}^{e}+\left( \hat{\partial}%
_{d}f_{~[c}^{e}\right) f_{~n]}^{d}\right) \right) \hat{\partial}_{e}h
\end{eqnarray*}%
where $h$ is an arbitrary matrix-valued function of $x$ and $s$.

\noindent Condition $\left( \text{%
%TCIMACRO{\TeXButton{\ref{cond Qhat4}}{\ref{cond Qhat4}}}%
%BeginExpansion
\ref{cond Qhat4}%
%EndExpansion
}\right) $\ becomes:%
\begin{eqnarray*}
\left( \text{%
%TCIMACRO{\TeXButton{\ref{cond Qhat4}}{\ref{cond Qhat4}}}%
%BeginExpansion
\ref{cond Qhat4}%
%EndExpansion
}\right) &=&-R_{~m[ac}^{b}f_{~n]}^{m}+\nabla _{\lbrack n}\nabla
_{c}f_{~a]}^{b}+\left( \nabla _{\lbrack n}f_{~c}^{e}+\left( \hat{\partial}%
_{d}f_{~[c}^{e}\right) f_{~n}^{d}\right) \hat{\partial}_{|e|}f_{~a]}^{b} \\
&=&-R_{~m[ac}^{b}f_{~n]}^{m}+R_{~[anc]}^{d}f_{~d}^{b}-R_{~d[nc}^{b}f_{~a]}^{d}+\left( 
\hat{\partial}_{d}f_{~[a}^{b}\right) R_{nc]~e}^{~~~d}s^{e} \\
&&+\left( \nabla _{\lbrack n}f_{~c}^{e}+\left( \hat{\partial}%
_{d}f_{~[c}^{e}\right) f_{~n}^{d}\right) \hat{\partial}_{|e|}f_{~a]}^{b} \\
&=&\left( \hat{\partial}_{d}f_{~[a}^{b}\right) R_{nc]~e}^{~~d}s^{e}+\left(
\nabla _{\lbrack n}f_{~c}^{d}+\left( \hat{\partial}_{e}f_{~[c}^{d}\right)
f_{~n}^{e}\right) \hat{\partial}_{|d|}f_{~a]}^{b} \\
&=&\hat{\partial}_{d}f_{~[a}^{b}\left( R_{nc]~e}^{~~d}s^{e}+\nabla
_{n}f_{~c]}^{d}+\left( \hat{\partial}_{|e|}f_{~c}^{d}\right)
f_{~n]}^{e}\right)
\end{eqnarray*}%
Now according to the original condition $\left( \text{%
%TCIMACRO{\TeXButton{\ref{cond Qhat}}{\ref{cond Qhat}}}%
%BeginExpansion
\ref{cond Qhat}%
%EndExpansion
}\right) $ becomes the equation:%
\begin{equation*}
\nabla _{\lbrack n}f_{~c]}^{d}+\left( \hat{\partial}_{e}f_{~[c}^{d}\right)
f_{~n]}^{e}=-R_{~enc}^{d}s^{e}
\end{equation*}%
therefore:%
\begin{equation*}
\left( \text{%
%TCIMACRO{\TeXButton{\ref{cond Qhat4}}{\ref{cond Qhat4}}}%
%BeginExpansion
\ref{cond Qhat4}%
%EndExpansion
}\right) =\hat{\partial}_{d}f_{~[a}^{b}\left(
R_{nc]~e}^{~~d}s^{e}-R_{~|e|nc]}^{d}s^{e}\right) =0
\end{equation*}%
so modulo the original condition $\left( \text{%
%TCIMACRO{\TeXButton{\ref{cond Qhat}}{\ref{cond Qhat}}}%
%BeginExpansion
\ref{cond Qhat}%
%EndExpansion
}\right) $ the local integrability condition $\left( \text{%
%TCIMACRO{\TeXButton{\ref{cond Qhat4}}{\ref{cond Qhat4}}}%
%BeginExpansion
\ref{cond Qhat4}%
%EndExpansion
}\right) $ is zero. Therefore $f_{~a}^{b}$\ exists at least locally by the
Cauchy-Kovalevskaya theorem.

\noindent \textbf{QED.}

\section{Appendix C: Change of Embedding}

\noindent 
%TCIMACRO{%
%\TeXButton{\hypertarget{Appendix C}{}}{\hypertarget{Appendix C}{}}}%
%BeginExpansion
\hypertarget{Appendix C}{}%
%EndExpansion
In this appendix we want to find $\hat{x}$ and $\hat{p}$ associated to the
embedding:%
\begin{equation*}
x^{\mu }x_{\mu }=1/C~~\ ,~~\ x^{\mu }p_{\mu }=A
\end{equation*}%
To do this we exploit the canonical transformation:%
\begin{equation*}
\tilde{p}_{\mu }=p_{\mu }+CAx_{\mu }~~~,~~~\tilde{x}^{\mu }=x^{\mu }
\end{equation*}%
Because it leaves the symplectic form invariant:%
\begin{equation*}
\tilde{\omega}=d\tilde{p}_{\mu }d\tilde{x}^{\mu }=dp_{\mu }dx^{\mu }=\omega
\end{equation*}%
as well as all other conditions except $x^{\mu }\tilde{p}_{\mu }=A$:%
\begin{equation*}
dx^{\mu }\tilde{p}_{\mu }+x^{\mu }d\tilde{p}_{\mu }=0
\end{equation*}%
it also leaves $D$ and $\hat{D}$ unchanged. Therefore the two solutions:%
\begin{equation*}
\hat{x}^{\mu }=\left( x^{\mu }+s^{\mu }\right) \frac{1}{\sqrt{Cu+1}}
\end{equation*}%
\begin{equation*}
\hat{p}_{\mu }=\left( -Cz_{\nu }s^{\nu }x_{\mu }+z_{\mu }\right) \sqrt{Cu+1}%
-iC\hbar n\hat{x}_{\mu }
\end{equation*}%
are solutions still.

We perform the canonical transformation:%
\begin{equation*}
\hat{x}^{\mu }=\left( x^{\mu }+s^{\mu }\right) \frac{1}{\sqrt{Cu+1}}
\end{equation*}%
\begin{equation*}
\hat{p}_{\mu }=\left( -Cz_{\nu }s^{\nu }x_{\mu }+z_{\mu }\right) \sqrt{Cu+1}%
-iC\hbar n\hat{x}_{\mu }
\end{equation*}%
where:%
\begin{equation*}
p_{\mu }=\tilde{p}_{\mu }-CAx_{\mu }
\end{equation*}%
and:%
\begin{equation*}
\widehat{\tilde{p}}_{\mu }=\hat{p}_{\mu }-CA\hat{x}_{\mu }
\end{equation*}%
therefore:%
\begin{equation*}
\hat{x}^{\mu }=\left( x^{\mu }+s^{\mu }\right) \frac{1}{\sqrt{Cu+1}}
\end{equation*}%
\begin{equation*}
\widehat{\tilde{p}}_{\mu }=\left( z_{\mu }-Cz_{\nu }s^{\nu }x_{\mu }\right) 
\sqrt{Cu+1}-C\left( i\hbar n+A\right) \hat{x}_{\mu }
\end{equation*}

\section{Appendix D: The Derivation of the Phase-Space Connection}

%TCIMACRO{%
%\TeXButton{\hypertarget{Appendix D}{}}{\hypertarget{Appendix D}{}}}%
%BeginExpansion
\hypertarget{Appendix D}{}%
%EndExpansion
Given the Levi-Civita connection $\nabla $ on the configuration space $M$
and subsequent curvature given the metric $g$ on a general manifold $M$:

\begin{equation}
\nabla _{\sigma }f\left( x\right) =\frac{\partial f}{\partial x^{\sigma }}
\end{equation}%
\begin{equation*}
\nabla _{\sigma }\left( dx^{\mu }\right) =-\Gamma _{~\nu \sigma }^{\mu
}dx^{\nu }
\end{equation*}%
\begin{equation*}
\nabla _{\sigma }\left( \frac{\partial }{\partial x^{\mu }}\right) =\Gamma
_{~\mu \sigma }^{\nu }\frac{\partial }{\partial x^{\nu }}
\end{equation*}%
\begin{equation*}
\nabla _{\lbrack \sigma }\nabla _{\rho ]}\left( dx^{\mu }\right) =R_{~\nu
\sigma \rho }^{\mu }dx^{\nu }
\end{equation*}%
where $R_{~\nu \sigma \rho }^{\mu }$ is the Riemann tensor. Of course we
have the conditions that $\nabla $ preserves the metric $g$ and is
torsion-free:%
\begin{equation*}
\nabla _{a}g_{bc}=0
\end{equation*}%
\begin{equation*}
\nabla _{\lbrack a}\nabla _{b]}f\left( x\right) =0
\end{equation*}%
for all functions $f\left( x\right) $. Together these uniquely fix $\nabla $
and give the standard formula for the Christoffel symbols:%
\begin{equation}
\Gamma _{~\mu \nu }^{\rho }=-\frac{1}{2}g^{\rho \sigma }\left( \partial
_{\mu }g_{\nu \sigma }+\partial _{\nu }g_{\mu \sigma }-\partial _{\sigma
}g_{\mu \nu }\right)  \label{Gamma}
\end{equation}%
where $\partial _{\mu }$ are the partial derivatives in some basis $x^{\mu }$%
.Define now a basis of covectors or forms $\Theta ^{B}\in T^{\ast }T^{\ast
}M $ (the cotangent bundle of the phase-space):%
\begin{equation*}
\Theta ^{B}=\left( dx^{\sigma },\alpha _{\sigma }\right)
\end{equation*}%
where the $dx$'s are the first $n$ $\Theta $'s, the $\alpha $'s are the last 
$n$ $\Theta $'s and they are defined to be:%
\begin{equation}
\alpha _{\mu }:=dp_{\mu }-\Gamma _{~\mu \rho }^{\nu }dx^{\rho }p_{\nu }
\end{equation}%
To extend $D$ to define $D\otimes \alpha _{\mu }$\ we require that $D$
preserves the symplectic form $\omega $:%
\begin{equation*}
0=D\otimes \omega =D\otimes \left( \alpha _{\mu }dx^{\mu }\right) \implies
\left( D\otimes \alpha _{\mu }\right) dx^{\mu }=\left( \Gamma _{~\mu \sigma
}^{\nu }dx^{\sigma }\otimes \alpha _{\nu }\right) dx^{\mu }
\end{equation*}%
where it can be shown that:%
\begin{equation}
\omega =\alpha _{\mu }dx^{\mu }=dp_{\mu }dx^{\mu }  \label{omega}
\end{equation}%
which can be proven by the torsion-free condition which tells us that $%
\Gamma _{~[\mu \rho ]}^{\nu }=0$. Therefore we make the ansatz:%
\begin{equation}
D\otimes \alpha _{\mu }:=S_{\mu \rho \sigma }dx^{\sigma }\otimes dx^{\rho
}+\Gamma _{~\mu \sigma }^{\nu }dx^{\sigma }\otimes \alpha _{\nu }
\label{Dalpha1}
\end{equation}%
where $S_{[\mu \rho ]\sigma }=0$.

We can fix $S_{\mu \rho \sigma }$ by requiring that the directional
derivative $\mathcal{D}_{v}$ of a vector and covector in any direction $%
v^{a} $ on the manifold is also a vector and covector respectively.%
\begin{equation*}
w_{\mu }\text{ is a covector }\iff \mathcal{D}_{v}w_{\mu }:=v^{\rho }\left(
\partial _{\rho }w_{\mu }-\Gamma _{~\mu \rho }^{\nu }w_{\nu }\right) \text{
is a covector}
\end{equation*}%
\begin{equation*}
w^{\mu }\text{ is a vector }\iff \mathcal{D}_{v}w^{\mu }:=v^{\rho }\left(
\partial _{\rho }w^{\mu }+\Gamma _{~\nu \rho }^{\mu }w^{\nu }\right) \text{
is a vector}
\end{equation*}%
this means that for any $p_{\mu }=w_{\mu }\left( x\right) $ (i.e., any
section in the cotangent bundle) the directional derivative of a covector is
a covector. Then the following formula must hold:%
\begin{equation*}
\nabla _{\lbrack a}\nabla _{b]}w_{c}=R_{~cab}^{d}w_{d}
\end{equation*}%
for every $w_{\mu }$ by the definition of the Riemann tensor. This formula
then fixes the skew part of equation $\left( \text{%
%TCIMACRO{\TeXButton{\ref{Dalpha1}}{\ref{Dalpha1}}}%
%BeginExpansion
\ref{Dalpha1}%
%EndExpansion
}\right) $ to be:%
\begin{equation*}
D\alpha _{\mu }:=D\wedge \alpha _{\mu }=S_{\mu \rho \sigma }dx^{\sigma
}dx^{\rho }+\Gamma _{~\mu \sigma }^{\nu }dx^{\sigma }\alpha _{\nu }=-R_{~\mu
\rho \sigma }^{\nu }p_{\nu }dx^{\sigma }dx^{\rho }+\Gamma _{~\mu \sigma
}^{\nu }dx^{\sigma }\alpha _{\nu }
\end{equation*}%
\begin{equation*}
\implies S_{a[ce]}=-R_{~ace}^{b}p_{b}
\end{equation*}%
Therefore we need to solve for $S_{ace}$ that satisfies the two conditions:%
\begin{equation*}
S_{[ac]e}=0~~~\&~~~S_{a[ce]}=-R_{~ace}^{b}p_{b}
\end{equation*}%
Let $S_{ace}:=S_{~ace}^{b}p_{b}$ and these conditions become:%
\begin{equation}
S_{~[ac]e}^{b}=0~~~\&~~~S_{~a[ce]}^{b}=-R_{~ace}^{b}  \label{SR_eqn}
\end{equation}%
Using the first Bianchi identity, the solution to this equation is:%
\begin{equation}
S_{~ace}^{b}=-\frac{4}{3}R_{~(ac)e}^{b}  \label{SR_soln}
\end{equation}%
Therefore:%
\begin{equation*}
D\otimes \alpha _{\mu }:=-\frac{4}{3}R_{~(\mu \sigma )\beta }^{\psi }p_{\psi
}dx^{\beta }\otimes dx^{\sigma }+\Gamma _{~\mu \sigma }^{\nu }dx^{\sigma
}\otimes \alpha _{\nu }
\end{equation*}%
The phase-space connection is:%
\begin{equation}
Dx^{\mu }:=dx^{\mu }
\end{equation}%
\begin{equation*}
Dp_{\mu }:=dp_{\mu }
\end{equation*}%
\begin{equation*}
D\otimes dx^{\mu }=-\Gamma _{~\sigma \nu }^{\mu }dx^{\nu }\otimes dx^{\sigma
}
\end{equation*}%
\begin{equation*}
D\otimes \alpha _{\mu }=\Theta ^{B}\otimes D_{B}\alpha _{\mu }:=-\frac{4}{3}%
R_{~(\mu \sigma )\beta }^{\psi }p_{\psi }dx^{\beta }\otimes dx^{\sigma
}+\Gamma _{~\mu \sigma }^{\nu }dx^{\sigma }\otimes \alpha _{\nu }
\end{equation*}%
\begin{equation*}
\alpha _{\mu }:=dp_{\mu }-\Gamma _{~\mu \rho }^{\nu }dx^{\rho }p_{\nu }
\end{equation*}%
which is the connection in $\left( \text{%
%TCIMACRO{\TeXButton{\ref{D}}{\ref{D}}}%
%BeginExpansion
\ref{D}%
%EndExpansion
}\right) $ and the corresponding curvature:%
\begin{equation}
D^{2}x^{\mu }=0
\end{equation}%
\begin{equation*}
D^{2}p_{\mu }=0
\end{equation*}%
\begin{equation}
D^{2}\otimes dx^{\mu }=dx^{\sigma }dx^{\rho }\otimes R_{~\nu \sigma \rho
}^{\mu }dx^{\nu }  \notag
\end{equation}%
\begin{equation}
D^{2}\otimes \alpha _{\mu }=\frac{4}{3}dx^{\sigma }\left( C_{~\mu \beta \nu
\sigma }^{\psi }p_{\psi }dx^{\nu }+R_{~(\mu \beta )\sigma }^{\nu }\alpha
_{\nu }\right) \otimes dx^{\beta }-R_{~\mu \sigma \beta }^{\nu }dx^{\sigma
}dx^{\beta }\otimes \alpha _{\nu }  \notag
\end{equation}%
which is the curvature in $\left( \text{%
%TCIMACRO{\TeXButton{\ref{D2}}{\ref{D2}}}%
%BeginExpansion
\ref{D2}%
%EndExpansion
}\right) $ where $C_{~abes}^{c}:=\nabla _{s}R_{~(ab)e}^{c}$ and according to 
$\left( \text{%
%TCIMACRO{\TeXButton{\ref{nabla}}{\ref{nabla}}}%
%BeginExpansion
\ref{nabla}%
%EndExpansion
}\right) $ the formula for the curvature is:%
\begin{equation}
R_{~\nu \sigma \rho }^{\mu }=-\partial _{\lbrack \sigma }\Gamma _{~\rho ]\nu
}^{\mu }+\Gamma _{~\nu \lbrack \sigma }^{\kappa }\Gamma _{~\rho ]\kappa
}^{\mu }
\end{equation}%
We can extend to higher order tensors by using the Leibnitz rule and the
fact that $D$ and $\nabla $ commute with contractions.

\section{References}

\noindent 
%TCIMACRO{\TeXButton{\hypertarget{ref1}{}}{\hypertarget{ref1}{}}}%
%BeginExpansion
\hypertarget{ref1}{}%
%EndExpansion
Barton G. 1963 \textit{Introduction to Advanced Field Theory} (New York:
Interscience).

\noindent Bayen F. \textit{et al} 1978 \textit{Ann. Physics} \textbf{111},
61.

\noindent Birrell N. and Davies P. 1982 \textit{Quantum Fields in Curved
Space}, Cambridge: Cambridge University Press.

\noindent Bordemann M. \textit{et al} 2003 \textit{J. Funct. Anal. }\textbf{%
199} (1), 1 \textit{Preprint}\ 
%TCIMACRO{%
%\TeXButton{math.QA/9811055}{\href{http://xxx.lanl.gov/abs/math.QA/9811055}{math.QA/9811055}}}%
%BeginExpansion
\href{http://xxx.lanl.gov/abs/math.QA/9811055}{math.QA/9811055}%
%EndExpansion
.

\noindent Bordemann M. \textit{et al} 1998 \textit{Lett. Math. Phys.} 
\textbf{45}, 49.

\noindent Bordemann M. and Waldmann S. 1998 \textit{Commun. Math. Phys.} 
\textbf{195}, 549.

\noindent Connes A. \textit{et al} 1992 \textit{Lett. Math. Phys.} \textbf{24%
}, 1.

\noindent Dito G. 2002 \textit{Proc. Int. Conf. of 68}$^{\text{\textit{\`{e}%
me}}}$\textit{\ Rencontre entre Physiciens Th\'{e}oriciens et Math\'{e}%
maticiens on Deformation Quantization I (Strasbourg)}, (Berlin: de Gruyter)
p 55 \textit{Preprint} 
%TCIMACRO{%
%\TeXButton{math.QA/0202271}{\href{http://xxx.lanl.gov/abs/math.QA/0202271}{math.QA/0202271}}}%
%BeginExpansion
\href{http://xxx.lanl.gov/abs/math.QA/0202271}{math.QA/0202271}%
%EndExpansion
.

\noindent Dito G. and Sternheimer D. 2002 \textit{Proc. Int. Conf. of 68}$^{%
\text{\textit{\`{e}me}}}$\textit{\ Rencontre entre Physiciens Th\'{e}%
oriciens et Math\'{e}maticiens on Deformation Quantization I (Strasbourg)},
(Berlin: de Gruyter) p 9 \textit{Preprint} 
%TCIMACRO{%
%\TeXButton{math.QA/0201168}{\href{http://xxx.lanl.gov/abs/math.QA/0201168}{math.QA/0201168}}}%
%BeginExpansion
\href{http://xxx.lanl.gov/abs/math.QA/0201168}{math.QA/0201168}%
%EndExpansion
.

\noindent Fedosov\ B. 1996\textit{\ Deformation Quantization and Index Theory%
} (Berlin: Akademie).

\noindent Fr\o nsdal C. 1965 \textit{Rev. Mod. Phys.} \textbf{37}, 221.

\noindent Fr\o nsdal C. 1973 \textit{Phys. Rev.} D \textbf{10}, 2, 589.

\noindent Fr\o nsdal C. 1975a \textit{Phys. Rev.} D \textbf{12}, 12, 3810.

\noindent Fr\o nsdal C. 1975b \textit{Phys. Rev.} D \textbf{12}, 12, 3819.

\noindent Gerstenhaber M. 1964 \textit{Ann. Math.} \textbf{79}, 59.

\noindent Groenewold H. 1946 \textit{Physica} \textbf{12}, 405.

\noindent Gadella M. \textit{et al\ }2005\textit{\ J. Geom. Phys.} \textbf{55%
}, 316.

\noindent Hancock\ J. \textit{et al} 2004 \textit{Eur. J. Phys.} \textbf{25,}
525.

\noindent Hirshfeld A. 2003 \textit{Proc. IV Int. Conf. on Geometry,
Integrability and Quantization (Varna, Bulgaria)} (Coral Press, Sofia), p 11.

\noindent Hirshfeld A. and Henselder P. 2002a \textit{Am. J. Phys.} \textbf{%
70} (5), 537.

\noindent Hirshfeld A. and Henselder P. 2002b \textit{Annals Phys.} \textbf{%
298}, 382.

\noindent Hirshfeld A. and Henselder P. 2002c \textit{Annals Phys.} \textbf{%
302}, 59.

\noindent Hirshfeld A. and Henselder P. 2003 \textit{Annals Phys.} \textbf{%
308}, 311.

\noindent Hirshfeld A. \textit{et al} 2005 \textit{Annals Phys.} \textbf{317}%
, 107.

\noindent Hirshfeld A. \textit{et al} 2004 \textit{Annals Phys.} \textbf{314}%
, 75.

\noindent Kontsevich M. 2003 \textit{Lett. Math. Phys.} \textbf{66}, 157.

\noindent Moyal J. 1949 \textit{Proc. Cambridge Phil. Soc.} \textbf{45}, 99.

\noindent Tillman P. and Sparling G. 2006 \textit{J. Math. Phys.} \textbf{47}%
, 052102.

\noindent Tillman P. 2006a to appear in \textit{Proc. of II Int. Conf. on
Quantum Theories and Renormalization Group in Gravity and Cosmology},
Preprint 
%TCIMACRO{%
%\TeXButton{gr-qc/0610141}{\href{http://xxx.lanl.gov/abs/gr-qc/0610141}{gr-qc/0610141}}}%
%BeginExpansion
\href{http://xxx.lanl.gov/abs/gr-qc/0610141}{gr-qc/0610141}%
%EndExpansion
.

\noindent Tillman P. 2006b \textit{Deformation Quantization: From Quantum
Mechanics to Quantum Field Theory}, Preprint 
%TCIMACRO{%
%\TeXButton{gr-qc/0610159}{\href{http://xxx.lanl.gov/abs/gr-qc/0610159}{gr-qc/0610159}}}%
%BeginExpansion
\href{http://xxx.lanl.gov/abs/gr-qc/0610159}{gr-qc/0610159}%
%EndExpansion
.

\noindent Waldmann S. 2001 \textit{On the Representation Theory of
Deformation Quantization}, Preprint 
%TCIMACRO{%
%\TeXButton{math.QA/0107112}{\href{http://xxx.lanl.gov/abs/math.QA/0107112}{math.QA/0107112}}}%
%BeginExpansion
\href{http://xxx.lanl.gov/abs/math.QA/0107112}{math.QA/0107112}%
%EndExpansion
.

\noindent Waldmann S. 2004 \textit{States and Representations in Deformation
Quantization}, Preprint 
%TCIMACRO{%
%\TeXButton{math.QA/0408217}{\href{http://xxx.lanl.gov/abs/math.QA/0408217}{math.QA/0408217}}}%
%BeginExpansion
\href{http://xxx.lanl.gov/abs/math.QA/0408217}{math.QA/0408217}%
%EndExpansion
.

\noindent Wald R. 1984 \textit{General Relativity}, (Chicago: University of
Chicago Press).

\noindent Wald R. 1994 \textit{Quantum Field Theory in Curved Spacetime and
Black Hole Thermodynamics} (Chicago: University of Chicago Press).

\noindent Weinstein A. 1995 S\'{e}minaire Bourbaki, Vol. 1993/94, Ast\'{e}%
risque No. 227, Exp. No. 789, 5, 389-409.

\noindent Weinberg S. 1995 \textit{The Quantum Theory of Fields} (New York:
Cambridge University Press).

\noindent Weyl H. 1931 \textit{The Theory of Groups and Quantum Mechanics
(translated)}, (Dover: New York).

\noindent Wigner E. 1932 \textit{Phys. Rev.} \textbf{40}, 749.

\noindent Zachos C. and Curtright T. 1999 \textit{Prog. Theor. Phys. Suppl.} 
\textbf{135}, 244.

\noindent Zachos C. 2002 \textit{Int. J. Mod. Phys.} \textbf{A17}, 297.

\end{document}